\documentclass[journal]{IEEEtran}

 \usepackage[all=normal,paragraphs=tight,floats=normal,mathspacing=normal,wordspacing=tight,charwidths=tight,mathdisplays=normal,leading=normal]{savetrees}

\usepackage{amsmath,amsfonts}
\usepackage{amssymb}
\usepackage{array}
\usepackage[caption=false,font=normalsize,labelfont=sf,textfont=sf]{subfig}
\usepackage{textcomp}
\usepackage{stfloats}
\usepackage{url}
\usepackage{verbatim}
\usepackage{graphicx}
\usepackage{acronym}
\usepackage[bookmarks,colorlinks]{hyperref}
\usepackage{cite}
\usepackage{booktabs}
\usepackage{color,soul}
\usepackage{dsfont}
\usepackage{algorithm, algpseudocode}
\usepackage{tikz}
\usepackage{colortbl} 
\usepackage{multirow}
\usetikzlibrary{arrows.meta, positioning}

\def\BibTeX{{\rm B\kern-.05em{\sc i\kern-.025em b}\kern-.08em
    T\kern-.1667em\lower.7ex\hbox{E}\kern-.125emX}}
\usepackage{balance}

\usepackage{enumitem}

\definecolor{NewColor}{rgb}{0,0,0}
\definecolor{NewColor2}{rgb}{0,0,0}

\newcommand{\myVec}[1]{{\boldsymbol{#1}}}
\newcommand{\myMat}[1]{{\boldsymbol{#1}}}
\newcommand{\mySet}[1]{\mathcal{#1}}

\newcommand{\dnnParam}{\myVec{\theta}}
\newcommand{\desParam}{\hat{\myVec{\theta}}}
\newcommand{\bnnParam}{\myVec{\psi}}
\newcommand{\nnMap}{\myVec{h}}

\newcommand{\Nuser}{K}
\newcommand{\Nant}{N}

\acrodef{ai}[AI]{artificial intelligence}
\acrodef{dnn}[DNN]{deep neural network}
\acrodef{mse}[MSE]{mean squared error}
\acrodef{sgd}[SGD]{stochastic gradient descent}
\acrodef{mimo}[MIMO]{multiple-input multiple-output}
\acrodef{sic}[SIC]{soft interference cancellation}
\acrodef{mlp}[MLP]{multi-layer perceptron}
\acrodef{ekf}[EKF]{extended Kalman filter}
\acrodef{vi}[VI]{variational inference}
\acrodef{kld}[KLD]{Kullback–Leibler divergence}
\acrodef{elbo}[ELBO]{evidence lower bound}
\acrodef{vcl}[VCL]{Variational continual learning}
\acrodef{lf-vcl}[LF-VCL]{likelihood-focused VCL}
\acrodef{cl}[CL]{continual learning}
\acrodef{bnn}[BNN]{Bayesian neural network}
\acrodef{ssm}[SSM]{state space model}
\acrodef{snr}[SNR]{signal-to-noise ratio}
\acrodef{map}[MAP]{maximum {\em{a-posteriori}}}
\acrodef{csi}[CSI]{channel-state information}
\acrodef{llr}[LLR]{log-likelihood ratio}
\acrodef{ngd}[NGD]{natural gradient descent}
\acrodef{gd}[GD]{gradient descent}
\acrodef{ber}[BER]{bit error rate}
\acrodef{bbb}[BBB]{Bayes-by-backprop}
\acrodef{blr}[BLR]{Bayesian learning rule}
\acrodef{bog}[BOG]{Bayesian online gradient}
\acrodef{bong}[BONG]{Bayesian online natural gradient}
\acrodef{dag}[DAG]{directed acyclic graph}
\acrodef{fim}[FIM]{Fisher information matrix}
\acrodef{cm-ekf}[CM-EKF]{conditional-moments \ac{ekf}}
\acrodef{vd-ekf}[VD-EKF]{variational-diagonal extended Kalman filter}
\acrodef{ef}[EF]{empirical Fisher}
\acrodef{fc}[FC]{full covariance}
\acrodef{dg}[DG]{diagonal}
\acrodef{simd}[SIMD]{single-instruction multiple-data}
\acrodef{gpu}[GPU]{graphical processing unit}
\acrodef{ml}[ML]{maximum likelihood}
\acrodef{nlms}[NLMS]{normalized least mean squares}
\acrodef{oom}[OOM]{out of memory}
\acrodef{los}[LOS]{line-of-sight}
\acrodef{dlr}[DLR]{diagonal plus low-rank}
\acrodef{umi}[UMi]{urban microcell}
\acrodef{mmse}[MMSE]{minimum mean square error}
\acrodef{csi}[CSI]{channel state information}
\acrodef{maml}[MAML]{model-agnostic meta-learning}

\begin{document}
\title{Online Learning of Modular Bayesian Deep Receivers:\\Single-Step Adaptation with Streaming Data}
\author{Yakov Gusakov, Osvaldo Simeone, Tirza Routtenberg, and Nir Shlezinger
\thanks{Parts of this work were presented at the 2025 IEEE International Conference on Acoustics, Speech, and Signal Processing (ICASSP) as the paper~\cite{gusakov2025rapid}.
Y. Gusakov, T. Routtenberg, and N. Shlezinger are with the School of ECE, Ben-Gurion University of the Negev, Be'er-Sheva, Israel (e-mail: gusakovy@post.bgu.ac.il, \{tirzar; nirshl\}@bgu.ac.il).
{O. Simeone is with the Institute for Intelligent Networked Systems, Northeastern University London, One Portsoken Street, London, E1 8PH, UK (email: o.simeone@northeastern.edu).
This work was supported by the European Research Council (ERC) under the European Union’s Horizon Europe Programme under the ERC starting grant nr. 101163973 (FLAIR)  and  by the Israel Science Foundation (grant No. 3314/25). The work of O. Simeone was supported by the ERC under the European Union’s Horizon Europe Programme (grant agreement No. 101198347), by an Open Fellowship of the EPSRC (EP/W024101/1), and by the EPSRC project (EP/X011852/1).} }}

\maketitle

\begin{abstract}
\Ac{dnn}-based receivers offer a powerful alternative to classical model-based designs for wireless communication, especially in complex and nonlinear propagation environments. However, their adoption is challenged by the rapid variability of wireless channels, which makes pre-trained static \ac{dnn}-based receivers ineffective, and by the latency and computational burden of online \ac{sgd}-based learning.   
In this work, we propose an online learning framework that enables rapid low-complexity adaptation of \ac{dnn}-based receivers. Our approach is based on two main tenets. First, we cast online learning as Bayesian tracking in parameter space, enabling a single-step adaptation, which deviates from multi-epoch \ac{sgd}. Second, we focus on modular \ac{dnn} architectures that enable parallel, online, and localized variational Bayesian updates.   Simulations with practical communication channels demonstrate that our proposed online learning framework can maintain a low error rate with markedly reduced update latency and increased robustness to channel dynamics as compared to traditional gradient descent based method.
\end{abstract}

\acresetall

\section{Introduction}
\label{sec:intro}

\subsection{Context and Motivation}
\label{subsec:motive}
The increasing demand for wireless data motivates the integration of emerging technologies such as massive and holographic \ac{mimo}~\cite{huang2020holographic} and reconfigurable intelligent surfaces~\cite{liu2021reconfigurable}. While these innovations expand the capacity and functionality of next-generation networks, they also substantially complicate receiver design, challenging the applicability of classical model-based signal processing techniques~\cite{shlezinger2021dynamic, gabay2023wideband}. A promising direction to cope with such complexity is the incorporation of  \ac{ai} tools, and particularly \acp{dnn}, into the receiver chain~\cite{dai2020deep}. By learning to infer from data, deep receivers offer a model-agnostic alternative that can operate effectively in unknown and nonlinear  environments~\cite{oshea2017introduction, shlezinger2024artificial}.

Despite their promise, the deployment of deep receivers in practical wireless systems is subject to two fundamental constraints~\cite{tong2022nine}: 
$(i)$ wireless channels are inherently time-varying, implying that a suitable mapping learned by a \ac{dnn} during training may quickly become outdated; and $(ii)$ wireless devices typically operate under stringent constraints on computation, memory, and energy consumption. The first constraint motivates training deep receivers on-device, while the second constraint poses a challenge for standard deep learning algorithms, which are typically data- and compute-intensive~\cite{raviv2023adaptive}.

\begin{figure}
    \centering
    \includegraphics[width=\linewidth]{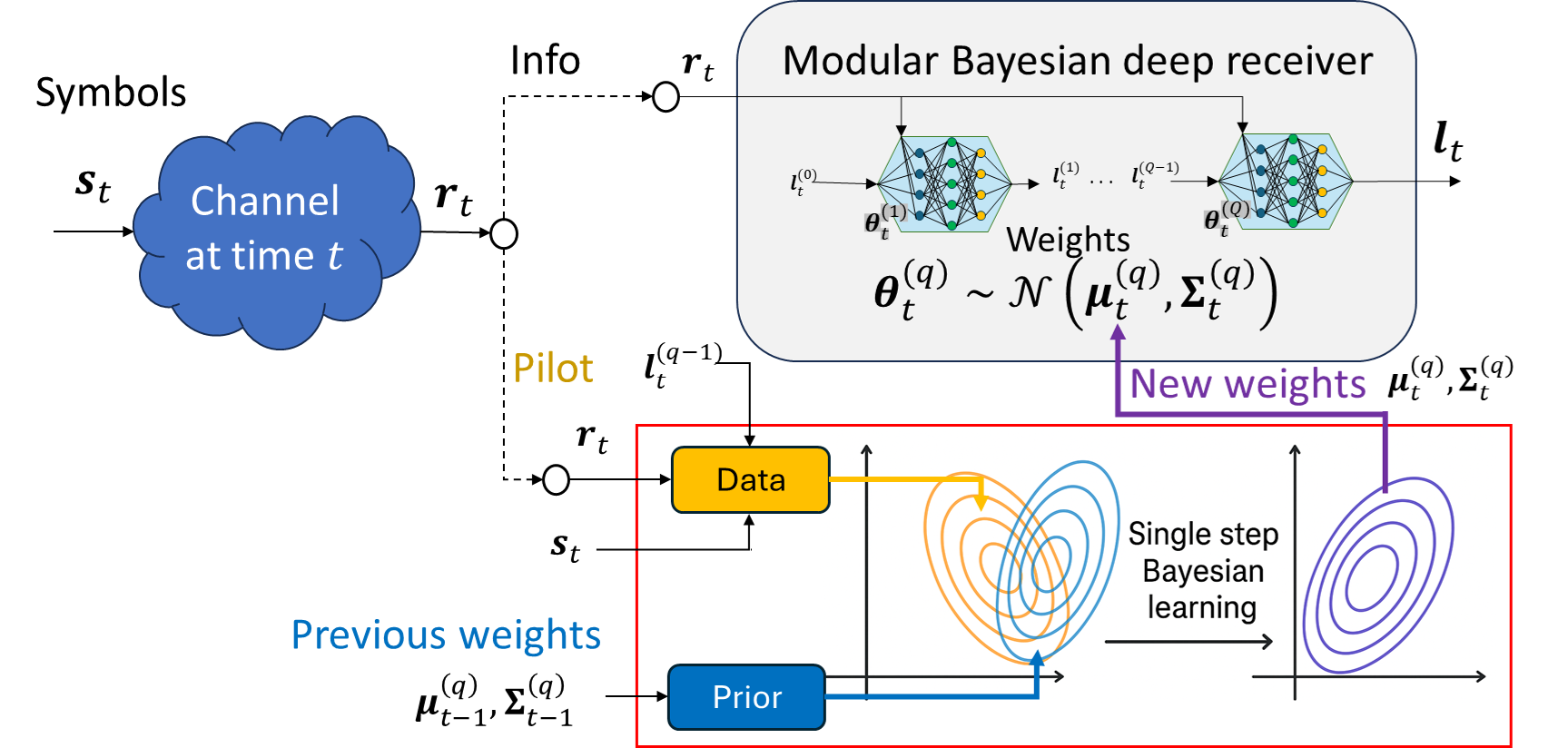}
    \caption{Illustration of proposed framework for single-step online adaptation of modular Bayesian deep receivers}
    \label{fig:illust}
\end{figure}

Two main paradigms are considered for incorporating deep learning into the physical layer of wireless receivers. The first relies on offline pre-trained \acp{dnn}, inheriting a common framework in traditional deep learning domains such as computer vision and natural language processing. These models are typically trained over large datasets encompassing diverse channel realizations, aiming to learn mappings that generalize across multiple environments via {\em joint learning}~\cite{oshea2017introduction,xia2020note}.  

Some level of adaptivity can be achieved with pre-trained models using {\em ensemble-based architectures}~\cite{raviv2024crc}, which maintain a set of specialized \acp{dnn} for different conditions; conventional linear channel estimates as additional inputs in inference~\cite{honkala2021deeprx,goutay2021machine}; or hypernetworks that alter the \ac{dnn} mapping based on an external \ac{dnn}~\cite{liu2024hypernetwork,raviv2024modular}. Despite their simplicity in deployment, these approaches tend to produce highly parameterized models with elevated memory and compute requirements, limiting their feasibility for edge devices. Moreover, such pre-trained receivers struggle to generalize to unseen channel distributions without explicit adaptation,  and often incur latency due to large inference pipelines\cite{Zecchin2024incontext}.

An alternative paradigm to using pre-trained models employs {\em online learning}, wherein the deep receiver is adapted continuously using data from the current operating conditions~\cite{raviv2023adaptive}. This enables adaptation to channel variations, maximizing performance by tailoring the receiver to the instantaneous channel distribution. However, online training with conventional deep learning methods is often impractical in wireless devices due to computational constraints and the lack of abundant labeled data. Real-time updates require significant processing power and memory, while pilots offer limited samples that may be insufficient for effective training.

In this work, we introduce a novel online learning paradigm for deep learning-aided wireless receivers, illustrated in Fig.~\ref{fig:illust}, which jointly leverages two complementary principles: $(i)$ modular designs of the receiver architecture; and $(ii)$ Bayesian \acp{dnn}. While modularity and Bayesian learning have each independently shown promise in facilitating efficient \ac{ai}-aided receivers, we combine them into a unified framework for {\em rapid adaptation} inspired by recent advances in continual Bayesian learning~\cite{Chang2022on,jones2024bayesian,duran2024unifying}.
The key conceptual insight of our approach is to cast the task of online training as the tracking of a dynamic system in the parameter space~\cite{durbin2012time}, where statistical observations stem from received pilot signals. This formulation enables the application of recursive Bayesian filtering methods to online learning, facilitating single-step updates in response to time-varying channel dynamics. 

\subsection{Related Work}
\label{subsec:relwork}
The complexity of online adaptation can be partially tackled by using several complementary approaches. These include architectures that integrate domain knowledge to yield compact and modular designs that reduce training complexity~\cite{shlezinger2019viterbinet,shlezinger2019deepsic,shlezinger2023model}; meta-learning techniques for optimizing the hyperparameters of the learning algorithm~\cite{park2020learning,raviv2022online}; and the incorporation of change detection mechanisms~\cite{uzlaner2024concept} to identify when retraining is necessary.

To obtain training data for online adaptation, it was proposed to exploit the structure of communication protocols~\cite{aoudia2021end, fischer2022adaptive} and modulation schemes~\cite{cohen2022symbol} for self-supervision. This approach can be combined with targeted data augmentation~\cite{huang2019data, raviv2022data} to obtain labeled data. Moreover,  Bayesian learning tools were proposed to avoid overfitting with limited data~\cite{simeone2022,cohen2022bayesian,zecchin2023robust}, and their improved calibration was shown to be highly effective when combined with modular deep architectures, enabling improved generalization and performance~\cite{RavivBayesian2024}.

Yet, a central bottleneck persists: nearly all existing methods rely on \ac{sgd} or its variants, which necessitate multiple passes over data and slow convergence. These characteristics make them ill-suited for scenarios where channel conditions evolve rapidly and adaptation must occur within the order of a coherence block of a wireless channel. This gives rise to the need for novel online learning mechanisms that can perform effective updates, by potentially deviating from conventional and widely-accepted \ac{sgd}-based learning.

\subsection{Main Contributions}
\label{ssec:contrib}
In this work, we propose a low-latency, high-performance online training algorithm for deep receivers based on Bayesian training and modular architectures. Our main contributions are summarized as follows:
\begin{itemize}
    \item \textbf{Online learning as Bayesian tracking:} We cast the online training of deep receivers as a Bayesian tracking task over a dynamic system defined by the evolving \ac{dnn} parameters via variational learning~\cite{farquhar2019unifying}. This enables the principled application of recursive Bayesian estimation tools,  which are well-suited for low-latency adaptation in dynamic environments.
    
    \item \textbf{Natural-gradient-based online Bayesian learning:} We adapt online Bayesian learning algorithms grounded in natural-gradient approximations to implement online learning via Bayesian tracking. 
    
    \item \textbf{Complexity-aware streaming modular adaptation:} We integrate the proposed Bayesian learning techniques with modular deep receivers to support \emph{module-wise} Bayesian updates, {and treat channel outputs as {\em streaming data} to enable Bayesian learning in a pipelined and asynchronous manner,} leading to scalable, fast, and localized online training with minimal overhead.
    
    \item \textbf{Extensive experimental validation:} We validate our proposed methodology through extensive simulations involving both synthetic and realistic wireless environments. In particular, we consider dynamic fading scenarios generated using the Sionna~\cite{Sionna}, COST2100~\cite{liu2012cost}, and QuaDRiGa~\cite{jaeckel2014quadriga} simulators. The experiments demonstrate our ability to rapidly adapt to time-varying conditions while achieving state-of-the-art performance in terms of uncoded \ac{ber} and robustness to channel variations.
\end{itemize}

The rest of this paper is structured as follows. Section~\ref{sec:sysmodel} presents the system model, comprised of both the channel model and receiver architecture. Section~\ref{sec:method} describes our modular online Bayesian learning approach, which we evaluate in  Section~\ref{sec:exp}, while Section~\ref{sec:conclusions} provides concluding remarks.

\subsection{Organization and Notation}
\label{ssec:notation}
Throughout this paper, we use boldface lower-case and upper-case letters for vectors (e.g., $\myVec{x}$) and matrices (e.g., $\myMat{X}$), respectively. The $(i,j)$th entry of  $\myMat{X}$ is denoted by $[\myMat{X}]_{i,j}$. Calligraphic letters denote sets, e.g., $\mySet{X}$, with $|\mySet{X}|$ being the cardinality of $\mySet{X}$. We use $\mathbb{R}$ and $\mathbb{C}$ for the sets of real and complex numbers, respectively. The notation $\mathbb{P}(\cdot)$ is the probability function, while $\mathbb{E}[\cdot]$ is the expectation operator. 
When a subscript is specified, e.g., 
$\mathbb{E}_{\dnnParam \sim q}[\cdot]$, 
the expectation is taken with respect to the random variable~$\dnnParam$ 
distributed according to $q(\dnnParam)$.
We use $(\cdot)^\top$, $\odot$, and $\text{diag}(\cdot)$ to denote the transpose operator, the element-wise product of two vectors, and a diagonal matrix with the elements of a vector on its diagonal, respectively.
\textcolor{NewColor}{We use $\Re\{\cdot\}$ and $\Im\{\cdot\}$ to denote the real and complex parts, respectively, of a scalar or vector.}
The gradient operator $\nabla_{\myVec{z}} f(\myVec{z})$
denotes a column vector of partial derivatives, i.e.,
$\nabla_{\myVec{z}} f = [ \frac{\partial f}{\partial z_1},
\ldots, \frac{\partial f}{\partial z_P} ]^{\!\top}$.

\section{System Model and Preliminaries}
\label{sec:sysmodel}
In this section, we introduce the system model, describing the time-varying channel model and transmission scheme in Subsection~\ref{subsec:channel}, and the modular receiver architecture in Subsection~\ref{subsec:transmit}. In Subsection~\ref{subsec:problem}, we formulate the online adaptation problem, and review key preliminaries in Subsection~\ref{ssec:prelim}. The main symbols defined in the following are summarized in Table~\ref{tab:notation}.

\begin{table}
    \centering
    \caption{Summary of symbols}
    \label{tab:notation}
    \begin{tabular}{>{\bfseries}l l}
        \toprule
        Symbol & Description \\
        \midrule
        $t$ & Time index \\
        $K$ & Number of receiver users \\
        $N$ & Number of receiver antennas \\
        $\mathcal{S}\subset\mathbb{C}$ & Symbol constellation \\
        $B = \log_2\left|{\mathcal{S}}\right|$ & Number of bits per symbol \\
        $\myVec{b}_t\in \{0,1\}^{KB}$ & Bit vector \\ 
        $\myVec{s}_t\in \mathcal{S}^K$ & Transmitted symbol vector \\
        $\myVec{r}_t \in \mathbb{C}^N$ & Received signal \\
        $\myVec{\ell}_t \in \left[0,1\right]^{KB}$ & Soft bit-level estimates \\
        $\myVec{x}_t \in \mathbb{R}^{2N+KB}$ & DeepSIC module input \\
        $P$ & Number of receiver parameters \\
        $\dnnParam_t\in \mathbb{R}^P$ & Parameters of the deep receiver \\
        \bottomrule
    \end{tabular}
\end{table}


\subsection{Time-Varying Channel Model}
\label{subsec:channel}

\subsubsection{Channel Model}
We consider an uplink \ac{mimo} system with $\Nuser$ single-antenna users communicating with an $\Nant$-antenna receiver. 
Each symbol $s_t^{(k)} \in \mathcal{S}$ transmitted from user $k$ at time $t$ is drawn from a constellation of size $|\mathcal{S}| = 2^B$, where each symbol modulates $B$ bits. In particular, let  
\begin{equation}
    \myVec{b}_t^{(k)} = \left[b_{t,1}^{(k)}, \dots, b_{t,B}^{(k)}\right]^\top \in \{0,1\}^B, \quad  k = 1, \dots, K, 
\end{equation}
denote the binary vector, which is mapped to the symbol $s_t^{(k)}$. 
The bit vector transmitted by all $K$ users at time $t$ is denoted
\begin{equation}
    \myVec{b}_t = \left[\left(\myVec{b}_t^{(1)}\right)^\top, \left(\myVec{b}_t^{(2)}\right)^\top, \dots, \left(\myVec{b}_t^{(K)}\right)^\top\right]^\top \in \{0,1\}^{K B},
\end{equation}
and the corresponding complete symbol vector is defined as
\begin{equation}
    \myVec{s}_t = \left[s_t^{(1)},\dots,s_t^{(K)}\right]\in \mathcal{S}^K.
\end{equation}

The received signal, $\myVec{r}_t \in \mathbb{C}^{N}$, is characterized by a generic time-varying conditional distribution,
\begin{equation} \label{eq:channel_prob}
  \mathbb{P}\left(\myVec{r}_t=\myVec{r}| \myVec{s}_{t}=\myVec{s}\right),
\end{equation}
representing dynamic channel conditions.
Beyond this general input-output relationship, we do not impose a specific channel model, allowing for complex and highly nonlinear channels. In the following, $\mathbb{P}(\cdot)$ denotes the true distribution governing the data as in \eqref{eq:channel_prob}, while $p(\cdot)$ denotes an induced or assumed probability measure.

\subsubsection{Temporal Variations}

A common approach in the \ac{mimo} literature adopts a block-fading assumption, where the channel is treated as approximately constant over a coherence interval and assumed to change independently between successive blocks~\cite{hassibi2003much}. While this model simplifies receiver design, it does not fully capture the nature of real-world wireless channels, which often exhibit continuous variations over time, leading to effects such as channel aging within an assumed coherence block~\cite{ChannelAging}. In this work, we allow the channel to change within each block. It is only assumed that the conditional probability from \eqref{eq:channel_prob} evolves smoothly over time.

\subsubsection{Transmission Scheme}
\begin{figure}
    \centering
    \includegraphics[width=0.47\textwidth]{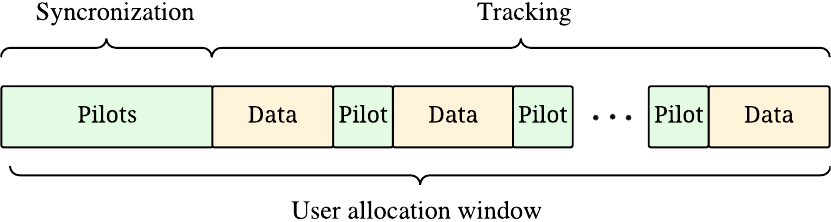}
    \caption{Transmission scheme with synchronization and  periodic pilot-data blocks.}
    \label{fig:Transmission Scheme}
\end{figure}

We consider a generic pilot-based transmission scheme as illustrated in Fig.~\ref{fig:Transmission Scheme}. Here, transmission consists of two phases:
\begin{enumerate}[leftmargin=*, label=\arabic*.]
    \item  A \emph{synchronization phase} during which $T_{\rm sync}$ predetermined  symbols $\left\{\myVec{s}_t\right\}_{t=1,\dots,T_{\rm sync}} $ are transmitted. 
    Such sequences, commonly employed by wireless protocols for transmission alignment, are used here for the initial calibration of our deep receiver.
    \item A \emph{tracking phase}, where data blocks are transmitted with pilot symbols periodically inserted between them. The receiver decodes the data symbols in real time, and uses the pilot symbols to fine-tune its operation, e.g., to adapt the deep receiver through online learning.
\end{enumerate}

\subsection{Receiver Architecture}
\label{subsec:transmit}

The demodulator is implemented as a \ac{dnn} parameterized by a vector of weights $\dnnParam_t$, which can be adapted over time to track changing channel conditions.  At each time step $t$, the \ac{dnn} maps the received signal $\myVec{r}_t \in \mathbb{C}^{N}$ to soft bit level estimates, denoted by
\begin{equation}
    \myVec{\ell}_t = \nnMap_{\dnnParam_t}\left(\myVec{r}_t\right) \in \left[0,1\right]^{K B}.
    \label{eq:receiver_llrs}
\end{equation}
The output vector is structured as
\begin{equation}
    \myVec{\ell}_t = \left[\ell_{t,1}^{(1)}, \dots, \ell_{t,B}^{(1)}, \dots, \ell_{t,1}^{(K)}, \dots, \ell_{t,B}^{(K)}\right]^\top,
    \label{eq:llr_vector}
\end{equation}
where $\ell_{t,i}^{(k)}$ is the soft estimate of the $i$-th bit of user~$k$ at time~$t$. It can be interpreted as the estimated conditional probability that $b_{t,i}^{(k)} = 1 $ conditioned on the received signal and the \ac{dnn} parameters, i.e., 
\begin{equation}
    \ell_{t,i}^{(k)} = {p(b_{t,i}^{(k)} = 1 \mid \myVec{r}_t, \dnnParam_t)}.
\end{equation}
For uncoded transmission, hard decisions on transmitted bits are made by thresholding the soft estimates, i.e., 
\begin{equation}
    \hat{\myVec{b}}_t = \mathds{1}[\myVec{\ell}_t > 0.5] \in \{0,1\}^{K B},
    \label{eq:receiver_symbol_decisions}
\end{equation}
where the comparison and the indicator function $\mathds{1}[\cdot]$ are applied element-wise.


\subsection{Problem Formulation}
\label{subsec:problem}
Our goal is to develop an online learning algorithm for enabling adaptive and flexible deep receivers by optimizing the \ac{dnn} parameters based on the incoming pilot symbols.
Formally, after receiving the channel output $\myVec{r}_t$ corresponding to a pilot symbol $\myVec{s}_t$, the algorithm maps them to an updated parameter vector $\dnnParam_t$. The receiver is then used to detect the subsequent data symbols.
The objective is to minimize the \ac{ber} in the data-bearing portion of the transmission. 

Let $\mathcal{T}_{\mathrm{data}}$ denote the set of time indices corresponding to the data symbols. The optimization problem can be expressed as:
\begin{equation}
    \min_{\{\dnnParam_t\}} \frac{1}{BK|\mathcal{T}_{\mathrm{data}}|} \sum_{t \in \mathcal{T}_{\mathrm{data}}} \sum_{k=1}^K \sum_{i=1}^B\mathbb{P}\left( \left. \hat{b}_{t,i}^{(k)} \neq b_{t,i}^{(k)} \right| \myVec{r}_t, \dnnParam_t \right),
\end{equation}
where the decoded bits $\hat{\myVec{b}}_t$ depend on $\dnnParam_t$ through the neural network mapping in~\eqref{eq:receiver_llrs} and~\eqref{eq:receiver_symbol_decisions}.

This optimization of the \ac{dnn} is to be carried out sequentially in an \emph{online} manner, updating $\dnnParam_t$ using only the information available up to time~$t$.
The online learning algorithm must be capable of tracking a smoothly time-varying, potentially nonlinear channel, while adhering to the practical constraints of wireless communication systems, namely, non-stationary signal statistics, and low-latency requirements. The key challenges include:
\begin{enumerate}[label={C\arabic*}]
    \item \label{itm:nonstat} \textbf{Non-stationarity:} The channel distribution evolves rapidly over time, requiring continuous adaptation of the receiver.
    \item \label{itm:latency} \textbf{Latency and complexity:} Receiver updates must be computationally efficient and meet strict runtime budgets for real-time operation.
    \item \label{itm:nonlin} \textbf{Nonlinearity:} The channel \eqref{eq:channel_prob} at each time $t$ may be highly nonlinear, difficult to parametrize explicitly, and possibly include non-Gaussian stochasticity. This makes classical model-based adaptation impractical.
\end{enumerate}


\subsection{Preliminaries}
\label{ssec:prelim}

\subsubsection{Bayesian Deep Learning}
\label{sssec:bnn}
Bayesian deep learning offers a framework to quantify uncertainty in \ac{dnn} predictions by treating the \ac{dnn} parameters $\dnnParam$ as random variables with prior $p\left(\dnnParam\right)$ and learning their posterior distribution $p(\dnnParam|\mySet{D})$ given the data $\mySet{D}$~\cite{wilson2020bayesian,simeone2022}. The posterior can be represented explicitly, as in variational inference~\cite{fortuin2022priors}, or implicitly via alternative induced trainable stochasticity, as in Monte Carlo dropout~\cite{gal2017concretedropout}.

Specifically, in standard (offline) variational inference, the posterior $p(\dnnParam|\mySet{D})$ is explicitly approximated by a \emph{variational distribution} $q_\bnnParam(\dnnParam)$, parametrized by the \emph{variational parameters} $\bnnParam$. These parameters are learned by the minimization of the \ac{elbo} objective, which consists of the expected negative log-likelihood regularized by a \ac{kld} term between the induced $q_\bnnParam(\dnnParam)$ and the prior $p(\dnnParam)$.
For a labeled dataset of the form $\mySet{D}=\{(\myVec{x}_i, \myVec{y}_i)\}_{i=1}^{|\mySet{D}|}$, this loss is given by
\begin{align}\mathcal{L}^{\text{ELBO}}_{\mySet{D}}\left(\bnnParam ; p(\dnnParam)\right) =& -\frac{1}{|\mySet{D}|}\sum_{i=1}^{|\mySet{D}|}\mathbb{E}_{\dnnParam \sim q_\bnnParam} \left[ \log p(\myVec{y}_i|\myVec{x}_i, \dnnParam) \right] \notag \\
    &+ D_{\text{KL}}\left(q_\bnnParam(\dnnParam) || p(\dnnParam)\right),
    \label{eq:ELBO}
\end{align}
where $D_{\mathrm{KL}}$ denotes the \ac{kld}. The \ac{elbo}  enables  gradient-based learning using, e.g., \ac{bbb}~\cite{blundell2015weight}.

During inference, $J$ realizations $\{\dnnParam_j\}_{j=1}^J$ are drawn in an i.i.d. manner from the distribution $q_\bnnParam$ to obtain an ensemble $\left\{p\left(\myVec{y}\mid \myVec{x},\dnnParam_j\right)\right\}_{j=1}^J$ whose predictive distributions are averaged to obtain a final prediction. Compared to standard (frequentist) \acp{dnn}, \acp{bnn} were shown to be less prone to overfitting, and their output probabilistic estimates are typically more calibrated, i.e., more faithful to the actual prediction uncertainty~\cite{gawlikowski2023survey}.

\subsubsection{Modular Deep Receivers}
\label{sssec:modular}

Modular deep receivers are deep learning-based architectures that incorporate task-specific structure by associating sub-networks with identifiable sub-tasks in the receiver processing chain. This design enables the use of compact \ac{dnn} modules with interpretable output features, localized parameter updates, and improved scalability~\cite{raviv2023adaptive}. In particular, such architectures naturally arise from model-based deep learning methods~\cite{shlezinger2022model}, where the algorithmic steps of classical receivers are integrated into the network structure via unfolding or architectural priors.

\begin{figure}
    \centering
    \includegraphics[width=0.47\textwidth]{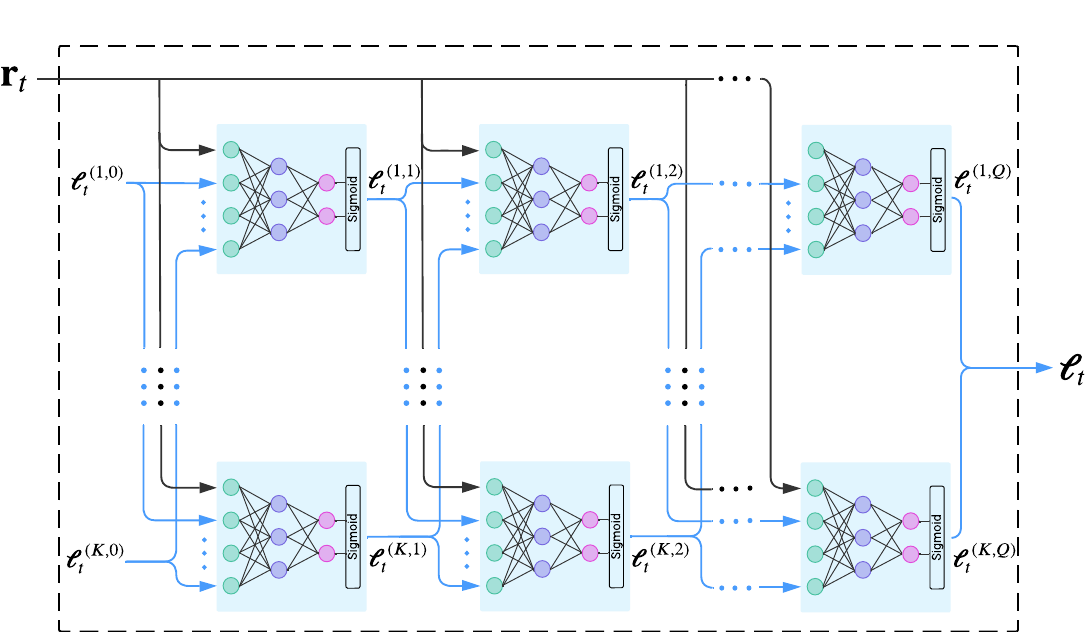}
    \caption{DeepSIC model architecture illustration.}
    \label{fig:deepsic}
\end{figure}

Our main example used in this paper of a modular deep receiver is \emph{DeepSIC}~\cite{shlezinger2019deepsic}, which unfolds the iterations of the \ac{sic} algorithm of \cite{choi2000iterative} for uplink \ac{mimo} detection into a modular \ac{dnn} architecture. As illustrated in Fig.~\ref{fig:deepsic}, DeepSIC consists of $Q$ iterations, each with $K$ separate modules.
The $k$th module at iteration $q$ is responsible for decoding the symbol of user $k$ while canceling the interference from the remaining users. The parameter vector of the receiver at time $t$, denoted $\dnnParam_t$, can be decomposed as
$ \dnnParam_t = \big\{\dnnParam_t^{(k,q)}\big\}_{k,q=1,1}^{K,Q}$,
where $\dnnParam_t^{(k,q)}$ denotes the weights of the $k$th user’s module at the $q$th \ac{sic} iteration.

Specifically, the $k$th module at iteration $q$ produces a vector of soft estimates, denoted by $\myVec{\ell}_{t}^{(k,q)}$, by applying
the user-specific \ac{dnn} mapping  $\nnMap_{\dnnParam_t^{(k,q)}}$ to an input vector shared by all modules in the layer, which is comprised of the received signals and the previous iteration's output as 
\textcolor{NewColor}{
\begin{equation} \label{eq:module_mapping}
    \myVec{x}_t^{(q)} = \left[\Re\left\{{\myVec{r}}_t^\top\right\}, \Im\left\{{\myVec{r}}_t^\top\right\}, \left(\myVec{\ell}_{t}^{(q-1)}\right)^\top\right]^\top,
\end{equation}
where
\begin{equation}
    \myVec{\ell}_{t}^{(q)} = \left[\left(\myVec{\ell}_{t}^{(1,q)}\right)^\top,\dots,\left(\myVec{\ell}_{t}^{(K,q)}\right)^\top\right]^\top.
\end{equation}
}
Consequently, the soft estimates are updated as
\begin{equation}
    \myVec{\ell}_{t}^{(k,q)} = \nnMap_{\dnnParam_t^{(k,q)}}\left( \myVec{x}_t^{(q)}\right).
\end{equation}
The initial soft estimates are  set to $\myVec{\ell}_{t}^{(k,0)} = \left[0.5,\dots,0.5\right]^\top$ for all $k=1,\dots,K$. The final soft estimates $\myVec{\ell}_t$ are taken from $\big\{\myVec{\ell}_{t}^{(k,Q)}\big\}_{k=1}^{K}$.

The resulting deep unfolded version of \ac{sic} enables accurate multi-user detection in complex nonlinear channels with non-Gaussian noise~\cite{shlezinger2019deepsic}, thus handling Challenge~\ref{itm:nonlin}. Moreover, the modular nature of DeepSIC was shown in \cite{RavivBayesian2024} to be beneficial when combined with \acp{bnn} for its internal modules, as the improved calibration of Bayesian deep learning leads to more accurate soft estimates exchanged by the modules, which in turn leads to improved performance.


{However, coping with nonstationarity (Challenge~\ref{itm:nonstat}) requires the receiver to adapt its parameters $\dnnParam_t$ (or their distribution parameters $\bnnParam_t$ for \acp{bnn}). To-date, online learning is still a complex and lengthy procedure (thus violating Challenge~\ref{itm:latency}), as regardless of whether the architecture is modular or black-box and whether one employs \acp{bnn} or frequentist \acp{dnn}, training is still carried out using numerous sequential gradient updates based on \ac{sgd} or its variants.
Still, in our current context of online learning, modular architectures, such as DeepSIC, offer several advantages: $(i)$ updates can be localized to specific modules corresponding to users with significant channel variation; $(ii)$ inference and training can be parallelized across modules; and $(iii)$ the architecture naturally supports a pipelined processing strategy. In our proposed methodology, we exploit these properties, combined with \acp{bnn}, to enable efficient and scalable online Bayesian learning via module-wise updates, as detailed in Section~\ref{sec:method}.}

\section{Modular Online Bayesian Learning}
\label{sec:method}
This section introduces our proposed methodology for rapid adaptation based on modular online Bayesian learning. 
The goal is to track time‑varying channels (Challenge~\ref{itm:nonstat}) using deep receivers capable of handling complex nonlinearities (Challenge~\ref{itm:nonlin}), while meeting stringent latency and complexity constraints (Challenge \ref{itm:latency}). 
In Subsection~\ref{subsec:ssm} we cast online training of \acp{bnn} as a {\em tracking} problem in a parameter‐space dynamical system, where the received pilots serve as streaming measurements. 
In Subsection~\ref{subsec:bayes} we leverage this state-space representation to perform online learning via efficient recursive Bayesian filtering.
This Bayesian framework, which integrates a level of uncertainty in the current model, enables replacing multiple gradient-descent iterations with a {\em single step} parameter update.
Then, in Subsection~\ref{subsec:Adaptation} we combine  Bayesian tracking with modular receiver architectures, and show how modularity facilitates low-complexity, low-latency online adaptation. We conclude by providing a discussion in Subsection~\ref{subsec:discuss}.

\subsection{State Space Formulation}
\label{subsec:ssm}
Our approach is based on modeling the online learning of deep receivers as a {\em dynamic system}. To formulate this, we consider a generic \ac{dnn}-based receiver, which at time $t$ is tasked with mapping its input $\myVec{x}_t$ into bit-wise soft estimates, $\myVec{\ell}_t = h_{\desParam_t}(\myVec{x}_t) \in (0,1)^B$, aiming to recover the true bits (labels) denoted $\myVec{b}_t \in \{0,1\}^B$. A special case of this generic model is a single DeepSIC module that can be obtained by omitting the superscript $k,q$.

We model channel variations in the \ac{dnn} parameter space via the {\em random process}, $\{\desParam_t\}_{t=1,2,\dots}$, where $\desParam_t$ denotes the desired \ac{dnn} parameters at time $t$. Considering relatively smooth variations in the desired \ac{dnn} parameters, we assume a first-order Gaussian Markov model of the form
\begin{equation} \label{eq:state_evo}
    p(\desParam_t | \desParam_{t-1} = {\myVec{\theta}}) = \mathcal{N}(\gamma \myVec{\theta}, \sigma^2 \myMat{I}),
\end{equation} 
where parameter $\sigma^2>0$ controls the evolution rate, while parameter $\gamma \in (0, 1]$ determines the memory of the process. Values near $1$ retain more past information, 
while smaller values enable faster adaptation to rapid changes. The initial weights are assumed to obey a Gaussian prior $p\big(\desParam_0\big) = \mathcal{N} \left(\myVec{\mu}_0, \myMat{\Sigma}_0\right)$.

At time $t$, the output of the \ac{dnn} receiver with the desired parameters $\desParam_t$, given by $\myVec{\ell}_t = h_{\desParam_t}(\myVec{x}_t)$,
dictates the distribution of the vector of independent Bernoulli variables, such that
\begin{equation} 
p(b_{t,i} = 1\mid \myVec{x}_t, \desParam_t) = \ell_{t,i}. \label{eq:likelihood}
\end{equation}
Consequently, the mean and covariance of the bit vector $\myVec{b}_t$ conditioned on the desired \ac{dnn} parameters $\desParam_t$ are given by
\begin{equation}
    \mathbb{E}\big[\myVec{b}_t | \myVec{x}_t, \desParam_t\big] \!=\! \nnMap_{\desParam_t}(\myVec{x}_t), \hspace{3.5cm}\label{eq:likelihood_mean} \end{equation}\begin{equation}
    \text{Cov}\big[\myVec{b}_t |  \myVec{x}_t, \desParam_t\big] \!=\! \text{diag}\big(\nnMap_{\desParam_t}(\myVec{x}_t) \odot (\myVec{1} \!- \!\nnMap_{\desParam_t}(\myVec{x}_t))\big).
    \label{eq:likelihood_cov}
\end{equation}
Equations \eqref{eq:state_evo}-\eqref{eq:likelihood} define a \emph{nonlinear \ac{ssm}} with \emph{Gaussian state evolution}, where the latent states are the desired \ac{dnn} parameters $\desParam_t \in \mathbb{R}^P$.


\subsection{Bayesian Online Learning}
\label{subsec:bayes}
We model online learning as the sequential estimation of desired \ac{dnn} parameters $\desParam_t$ evolving as a Markov process~\eqref{eq:state_evo} and generating noisy observations~\eqref{eq:likelihood} of the information bits.
Consequently, the online learning task is to process data points (pilots) that arrive sequentially and continuously update the belief about $\desParam_t$ as new pilot data $(\myVec{x}_t,\myVec{b}_t)$ arrives. To leverage this \ac{ssm} for learning purposes, we adopt a {\em Bayesian deep learning} perspective, and model the receiver as a \ac{bnn} whose parameters obey a Gaussian distribution (or a subfamily of Gaussian such as diagonal Gaussian). The variational distribution at time $t$, denoted by $q_{\bnnParam_t}\!\left(\dnnParam_t\right)$, is parameterized by a mean vector $\myVec{\mu}_t$ and a covariance matrix $\myMat{\Sigma}_t$, i.e., 
\begin{equation}
\label{eq:GaussBNN}
    q_{\bnnParam_t}(\desParam_t) = \mathcal{N}\left(\myVec{\mu}_t, \myMat{\Sigma}_t\right),
\end{equation}
which serves as a tractable approximation for the true posterior $p(\desParam_t \mid \{(\myVec{x}_\tau,\myVec{b}_\tau)\}_{\tau=1}^{t})$.


Casting the learning dynamics as a \ac{ssm} and the learning task as updating a Gaussian belief gives rise to two
online learning algorithms: 
$(i)$ an algorithm based on a variant of the \ac{ekf}~\cite{Tronarp2018CMEKF}, and $(ii)$ an algorithm based on combining Bayesian learning and natural gradient updates~\cite{jones2024bayesian}. 
Both approaches operate in a {\em one-shot} manner, allowing one to update with a {\em single gradient step} on a single data point. 

\subsubsection{Conditional-Moments EKF}

The \ac{ssm} representation in \eqref{eq:state_evo}--\eqref{eq:likelihood} suggests that online learning can be treated as the tracking of a latent state vector $\desParam_t$. Together with the imposed Gaussianity \eqref{eq:GaussBNN}, this naturally motivates the use of Kalman-type algorithms to recursively update the posterior.
Our primary learning method thus employs the \ac{cm-ekf} algorithm~\cite{Tronarp2018CMEKF}, which extends the \ac{ekf} algorithm to non-Gaussian \acp{ssm}. The \ac{cm-ekf} follows the standard \emph{predict}–\emph{update} structure of the \ac{ekf}. 

In the predict step, an approximate prior at time $t$ is obtained by propagating the previous posterior from time $t-1$ through the Gaussian state evolution model in~(\ref{eq:state_evo}). Given the previous posterior
$q_{\bnnParam_{t-1}}(\desParam_{t-1})$,
the predictive prior at time $t$ is given by:
\begin{equation}
\label{eq:int}
    q_{\bnnParam_{t \mid t-1}}(\desParam_t) = \int p(\desParam_t \mid \desParam_{t-1}) \, q_{\bnnParam_{t-1}}(\desParam_{t-1}) \, d\desParam_{t-1}.
\end{equation}
According to \eqref{eq:GaussBNN}, we assume that we have a Gaussian posterior, $q_{\bnnParam_{t-1}}(\desParam_{t-1}) = \mathcal{N}(\myVec{\mu}_{t-1}, \myMat{\Sigma}_{t-1})$. Thus, the integral in \eqref{eq:int} yields another Gaussian distribution \cite{papoulis2002prob} with mean and covariance, respectively, obtained as
\begin{align} 
    \myVec{\mu}_{t\mid t-1} &= \gamma \myVec{\mu}_{t-1}, \label{eq:predict_mean} \\ 
    \myMat{\Sigma}_{t\mid t-1} &=\gamma^2 \myMat{\Sigma}_{t-1} + \sigma^2 \myMat{I}. \label{eq:predict_cov}
\end{align}

In the update step, the predictive distribution is then used as a prior. 
Since the likelihood in \eqref{eq:likelihood} is a function of the network mapping, which is highly nonlinear, we employ a local linearization of the \ac{dnn} mapping that transforms \eqref{eq:likelihood_mean} into an affine transformation of the state $\desParam_t$ by taking its first-order Taylor series approximation around \eqref{eq:predict_mean}. Namely, we have
\begin{equation} \label{eq:linearize}
    \bar{\nnMap}_{\desParam_t} \left(\myVec{x}_t\right) = \nnMap_{\myVec{\mu}_{t\mid t-1}}\left(\myVec{x}_t\right)+ \myMat{H}_t \left(\desParam_t - \myVec{\mu}_{t\mid t-1}\right),
\end{equation}
where 
\[\myMat{H}_t = 
\left.\nabla_{\!\desParam}^{\!\top}\nnMap_{\desParam}(\myVec{x}_t)\right|_{\desParam=\myVec{\mu}_{t|t-1}}
\in \mathbb{R}^{KB\times P}\] is the Jacobian of the network with respect to its weights.

The variational posterior at time $t$ is obtained by the \ac{cm-ekf} update rule, which for our setting of online learning maps the \ac{dnn} input $\myVec{x}_t$ and the corresponding pilot bits $\myVec{b}_t$ into the \ac{bnn} parameters $\bnnParam_t$ via
\begin{align}
    \myVec{\mu}_t &= \myVec{\mu}_{t\mid t-1} + \myMat{K}_t\left(\myVec{b}_t-\nnMap_{\myVec{\mu}_{t|t-1}}\left(\myVec{x}_t\right)\right), \label{eq:ekf_mean_update}\\
    \myMat{\Sigma}_t &= \myMat{\Sigma}_{t\mid t-1} - \myMat{K}_t \myMat{H}_t\myMat{\Sigma}_{t\mid t-1}. \label{eq:ekf_cov_update}
\end{align}
Here $\myMat{K}_t$ is the Kalman gain matrix that is given by
\begin{equation}
    \!    \myMat{K}_t\! =\! \myMat{\Sigma}_{t| t\!-\!1} \myMat{H}_t^\top\left(\!\myMat{H}_t\myMat{\Sigma}_{t| t\!-\!1}\myMat{H}_t^\top+\myMat{R}_t\right)^{-1}\!\!,
\end{equation}
where 
\begin{equation} \label{eq:obs_cov}
    \myMat{R}_t\triangleq\text{Cov}\left[\myVec{b}_t\left| \myVec{x}_t,\desParam_t=\myVec{\mu}_{t\mid t-1}\right.\right]
\end{equation}
is computed according to~\eqref{eq:likelihood_cov} by setting $\desParam_t=\myVec{\mu}_{t\mid t-1}$. \textcolor{NewColor}{To ensure numerical stability, the diagonal entries of $\myMat{R}_t$ are clipped from below at a minimum value, which we set to $0.1$ in our numerical study.}

Online learning via the \ac{cm-ekf} updates the \ac{bnn} distribution parameters $(\myVec{\mu}_t, \myMat{\Sigma}_t)$ in a single step.
That is, each new observation $\myVec{x}_t$ is processed once by the network (to compute $\nnMap_{\myVec{\mu}_{t|t-1}}\left(\myVec{x}_t\right)$), and only a single backward pass is required (to compute $\myMat{H}_t$). This stands in contrast to standard \ac{sgd}-based learning, which typically requires numerous gradient steps. Moreover, \ac{cm-ekf} does not require sampling from the variational distribution, making it a robust deterministic learning rule.

While the complexity of each \ac{sgd} iteration only scales linearly with the number of \ac{dnn} parameters $P$, the complexity of processing one observation vector with the \ac{cm-ekf} is $\mathcal{O}\left(BP^2+B^3\right)$ where $BP^2$ is typically the dominant term, as $P\gg B$, i.e., the number of \ac{dnn} weights notably exceeds the number of bits extracted from $\myVec{x}_t$.

\subsubsection{Bayesian Online Natural Gradient}
\ac{cm-ekf} is essentially an algorithm for the tracking of a dynamic system, where the aim is to have the posterior mean estimate the latent state in the \ac{mse} sense~\cite{durbin2012time}. As the goal in online learning is not necessarily to minimize the \ac{mse} with respect to the assumed desired parameters, but rather to minimize some learning loss function, as in e.g., the \ac{elbo} \eqref{eq:ELBO}, one can formulate more conventional online learning algorithms that are derived from such loss formulations.  

Recent advances in Bayesian learning from streaming data have developed methods that enable single-step online learning based on the \ac{elbo}. In particular, the \ac{bong} algorithm proposed in \cite{jones2024bayesian} updates $\bnnParam_t$ based on the \ac{elbo} for the current sample $\left(\myVec{x}_t,\myVec{b}_t\right)$ by taking a single natural gradient step while using $\myVec{\mu}_{t|t-1}$ as a starting point and $q_{\bnnParam_{t\mid t-1}}$ as a prior (so that the KL term vanishes when evaluated at $\bnnParam=\bnnParam_{t|t-1}$).
The resulting online learning rule is given by  
\begin{equation} \label{eq:bong_update}
\!   \bnnParam_{t} \!=\! \bnnParam_{t| t\!-\!1}\! +\! \myMat{F}^{-1}_{\bnnParam_{t| t\!-\!1}} \nabla_{\bnnParam_{t| t\!-\!1}} \mathbb{E}_{\dnnParam \sim q_{\bnnParam_{t| t-1}}} \left[ \log p(\myVec{b}_t | \myVec{x}_t, \dnnParam) \right],
\end{equation}
where the \ac{fim} 
associated with the variational parameters $\bnnParam_{t|t-1}$, is defined as
\begin{eqnarray}
\label{eq:fisher_def}
    \myMat{F}_{\bnnParam_{t|t-1}} 
    = \hspace{6.7cm}\nonumber\\\mathbb{E}_{\dnnParam \sim q_{\bnnParam_{t|t-1}}}\hspace{-0.15cm}
      \left[ 
      \nabla_{\bnnParam_{t|t-1}} \hspace{-0.1cm}
      \log q_{\bnnParam_{t|t-1}}(\dnnParam) \,
      \nabla_{\bnnParam_{t|t-1}}^{\!\top}
     \hspace{-0.1cm} \log q_{\bnnParam_{t|t-1}}(\dnnParam)
      \right].
\end{eqnarray}

When employing a Gaussian variational distribution where the \ac{bnn} parameters are the natural parameters $\bnnParam_{t} = (\myVec{\Sigma}_t^{-1}\myVec{\mu}_t,  -\tfrac12\myVec{\Sigma}_t^{-1})$, then \eqref{eq:bong_update} reduces to two update equations~\cite{jones2024bayesian}, given by
\begin{align}
    &\myVec{\mu}_t = \myVec{\mu}_{t| t\!-\!1} \!+\! \myMat{\Sigma}_{t} {{\mathbb{E}_{\dnnParam_t\sim\bnnParam_{t| t\!-\!1}}\left[ \nabla_{\dnnParam_t} \log p\left(\myVec{b}_t\mid h_{\dnnParam_t}\left(\myVec{x}_t\right)\right)\right]}}, \label{eq:bong_mean_update}\\
    &\myMat{\Sigma}_t^{-1} = \myMat{\Sigma}_{t| t\!-\!1}^{-1} -  {{\mathbb{E}_{\dnnParam_t\sim\bnnParam_{t| t\!-\!1}}\left[ \nabla^2_{\dnnParam_t} \log p\left(\myVec{b}_t\mid h_{\dnnParam_t}\left(\myVec{x}_t\right)\right)\right]}}.
    \label{eq:bong_cov_update}
\end{align}

The \ac{bong} update rule~\eqref{eq:bong_mean_update}-\eqref{eq:bong_cov_update} requires computing  stochastic expectations. To translate it into an online learning method, we consider three different forms of approximations: 
\begin{enumerate}[label={A\arabic*}]
    
    \item \label{itm:LinGaus} {\bf Linearized Gaussian} approximations compute the expectations by adopting the linearized form~\eqref{eq:linearize} and imposing a Gaussian distribution on the likelihood, i.e., 
    \begin{equation} \label{eq:gauss_likelihood}
        p(\myVec{b}_t | \myVec{x}_t, \desParam_t) \approx \mathcal{N}\big(\bar{\nnMap}_{\desParam_t}(\myVec{x}_t), \myMat{R}_t\big),
    \end{equation}
    where $\myMat{R}_t$ is defined in~\eqref{eq:obs_cov}.

    \item \label{itm:EF} {\bf \Ac{ef}} replaces the Hessian in \eqref{eq:bong_cov_update} with the outer-product of the sampled gradients \cite{James2020Insights}.

    \item \label{itm:DLR} {\bf \Ac{dlr}} parameterization restricts the covariance in \eqref{eq:bong_mean_update} to be of the form~\cite{Mishkin2018Slang}
\begin{equation}
    \myMat{\Sigma}_t^{-1} = \myMat{D}_t + \myMat{W}_t\myMat{W}_t^\top,
    \label{eqn:DLR}
\end{equation}
where $\myMat{D}_t$ is a diagonal matrix and $\myMat{W}_t\in\mathbb{R}^{P\times R}$ with $R\ll P$.
\end{enumerate}

Among the above options, \ac{ef} approximations (\ref{itm:EF}) are typically less accurate and often lead to instability. 
When using a linearized Gaussian formulation (\ref{itm:LinGaus}), the stochastic expectations exhibit a closed-form expression of the model parameters and the linearized model $\myMat{H}_t$. In fact, under this modeling, the \ac{bong} update in~\eqref{eq:bong_mean_update}-\eqref{eq:bong_cov_update} coincides with the \ac{cm-ekf} update equations~\eqref{eq:ekf_mean_update}-\eqref{eq:ekf_cov_update}~\cite{jones2024bayesian}. While the \ac{dlr} parameterization (\ref{itm:DLR}) does not allow computing the expectations on its own, it can be combined with \ref{itm:LinGaus} to avoid the quadratic complexity growth of the \ac{cm-ekf} with the number of parameters $P$. Specifically, the \ac{bong} update under \ref{itm:LinGaus}+\ref{itm:DLR} is termed {\em \ac{vd-ekf}}~\cite{Chang2022on} when setting $R=0$ in  \eqref{eqn:DLR}, or {\em Lo-Fi}~\cite{Chang2023low} for $R>0$, and in both cases the resulting complexity only grows linearly with $P$. 

In summary, the \ac{bong} algorithm allows one-step online learning based on the \ac{elbo} objective. Although this step can be computationally heavy due to the need to evaluate stochastic expectations, it can be approximated with reduced complexity. The most stable and expressive approximation coincides with \ac{cm-ekf}, the fastest coincides with \ac{vd-ekf}, while Lo-Fi balances complexity and performance, as also shown in our numerical study in Section~\ref{sec:exp}.


\subsection{Modular Bayesian Online Adaptation}
\label{subsec:Adaptation}
The complexity associated with full covariance adaptation methods such as \ac{cm-ekf} grows at least quadratically with the number of \ac{dnn} weights $P$, notably limiting their usefulness for rapid adaptation of large \acp{dnn}. We next demonstrate how both complexity and latency can be alleviated by exploiting modular architectures such as DeepSIC. In the following, we build upon the state space formulation (Subsection~\ref{subsec:ssm}) and recursive online Bayesian learning algorithms (Subsection~\ref{subsec:bayes}), to show how modular architectures (focusing on DeepSIC) enable rapid online adaptation via module-wise updates, parallelization across user modules, and low-latency pipelining, harnessing the treatment of channel outputs as streaming data.

\begin{figure}
    \centering
    \includegraphics[width=0.47\textwidth]{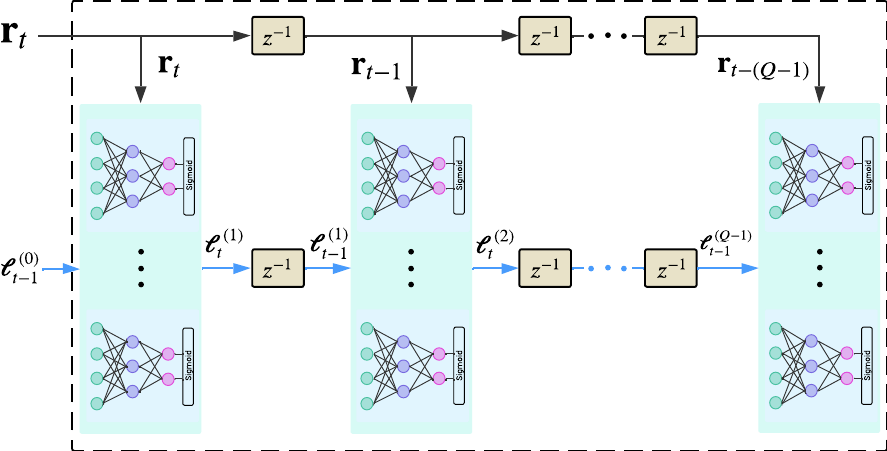}
    \caption{\textcolor{NewColor}{Pipelined DeepSIC model architecture illustration.}}
    \label{fig:pipeline}
\end{figure} 

{\bf Module-Wise Updates:} 
Each DeepSIC module, indexed by $(k,q)$, maintains its own Gaussian posterior $q_{\bnnParam_t}^{(k,q)}=\mathcal{N}(\myVec{\mu}_t^{(k,q)},\myMat{\Sigma}_t^{(k,q)})$. Since each module is tasked with producing a soft estimate of the $k$th user bits $\myVec{b}_t^{(k)}$, we can assign a label for each module, and thus formulate a {\em separate module-wise \ac{ssm}}. Accordingly, each module can be evaluated and trained separately via one-shot Bayesian online learning based on its corresponding \ac{ssm}. 

While \acp{bnn} typically come at the cost of slow inference due to the need to sample multiple i.i.d. realizations of $\dnnParam_t^{(k,q)}\sim q_{\bnnParam_t}^{(k,q)}$, here we employ the Bayesian modeling only for learning purposes. During inference, we plug in the mean parameter of the variational distribution, i.e., set $\dnnParam_t^{(k,q)}=\myVec{\mu}_t^{(k,q)}$, to avoid the overhead and excessive latency of sampling.

{\bf Parallelization:}
The ability to disentangle the update of each module can notably facilitate online learning, particularly when combined with some form of parallel processing capabilities. Specifically, all $K$ modules in iteration $q$ share the same input $\myVec{x}_t^{(q)}$ and can compute their update rules in parallel. In practice, this leverages either multicore or vectorization hardware to achieve up to $K$-fold speed-up over sequential processing.

{\bf Pipelining:}
Since DeepSIC iterates $q=1,2,\dots,Q$ sequentially, a conventional implementation would process each sample through all $Q$ iterations before processing the next. In our case, we view the channel outputs as streaming data, where both inference and online training are done on each sample rather than on aggregated batches. This approach allows us to pipeline $Q$ samples through the $Q$ iterations, so that at each time step, iteration $q$ processes sample $t-(q-1)$ for online adaptation or inference, as illustrated in Fig.~\ref{fig:pipeline}. This reduces overhead by up to a $Q$-fold, introducing minimal latency: at time $t$, the network outputs soft estimates of bits at $t-Q+1$. Such latency is often acceptable, particularly since digital communications typically employ subsequent channel coding, where a sequence of soft estimates are gathered at the output of the demodulator before the decoded bits are extracted, and thus there is some tolerable latency before data is decoded and processed.

{\bf Algorithm Summary:} 
\textcolor{NewColor}{Upon receiving pilot $(\myVec{r}_t,\myVec{b}_t)$, each module applies predict and update steps using $(\acute{\myVec{x}}_t^{(q)},\myVec{b}_{t-(q-1)}^{(k)})$, where 
\begin{equation}
    \acute{\myVec{x}}_t^{(q)} = \left[\Re\left\{{\myVec{r}}_{t-(q-1)}^\top\right\}, \Im\left\{{\myVec{r}}_{t-(q-1)}^\top\right\}, \left(\myVec{\ell}_{t-1}^{(q-1)}\right)^\top\right]^\top.
\end{equation}
}
The update step can be based on the \ac{cm-ekf}, or on other reduced complexity variants of \ac{bong} that employ \ac{dlr} approximations.
We summarize the online training process in Algorithm~\ref{alg:modular_update}, which fuses module-wise parallel filtering with pipelined execution.

\begin{algorithm}
\caption{Modular online Bayesian training at time $t$}
\label{alg:modular_update}
\begin{algorithmic}[1]
    \Require{Pilot queue $\{(\myVec{r}_{t-i},\myVec{b}_{t-i})\}_{i=0}^{Q-1}$}
    \smallskip
    \Require{\textcolor{NewColor}{Variational parameters $\big\{\bnnParam_{t-1}^{(k,q)}\big\}_{k,q=1,1}^{K,Q}$}}
    \smallskip
    \Require{\textcolor{NewColor}{Predictions from previous step $\{\myVec{\ell}_{t-1}^{(q)\top}\}_{q=1}^{Q-1}$}}
    \smallskip
    \State \textcolor{NewColor}{$\myVec{\ell}_{t-1}^{(0)} \gets [0.5,\dots,0.5]^\top$}
    \For{$q=1,\dots,Q$}
        \State $\tau \gets t-(q-1)$
        \State \textcolor{NewColor}{$\acute{\myVec{x}}_t^{(q)} \gets \left[\Re\left\{{\myVec{r}}_\tau^\top\right\}, \Im\left\{{\myVec{r}}_\tau^\top\right\}, \left(\myVec{\ell}_{t-1}^{(q-1)}\right)^\top\right]^\top$}
    \smallskip
        \ForAll{$k=1,\dots,K$}
            \State Predict: $\bnnParam_{t|t-1}^{(k,q)} \gets \text{predict}(\bnnParam_{t-1}^{(k,q)})$
            \State Update: $\bnnParam_{t}^{(k,q)} \gets \text{update}(\bnnParam_{t|t-1}^{(k,q)}, \acute{\myVec{x}}_t^{(q)}, \myVec{b}_\tau^{(k)})$
            \State Inference: $\myVec{\ell}_t^{(k,q)} \gets \nnMap_{\myVec{\mu}_t^{(k,q)}}(\acute{\myVec{x}}_t^{q})$
        \EndFor
        \State \textcolor{NewColor}{$\myVec{\ell}_{t}^{(q)\top} \gets \big[\big(\myVec{\ell}_{t}^{(1,q)}\big)^\top,\dots,\big(\myVec{\ell}_{t}^{(K,q)}\big)^\top\big]^\top$}
        \smallskip
    \EndFor
    \Ensure{\textcolor{NewColor}{Updated variational parameters $\big\{\bnnParam_{t}^{(k,q)}\big\}_{k,q=1,1}^{K,Q}$}}
\end{algorithmic}
\end{algorithm}

In terms of overhead, it is noted that each module-wise operation, i.e., the {\em Predict-Update-Inference} steps in Algorithm~\ref{alg:modular_update}, requires two forward passes of $\acute{\myVec{x}}_{t}^{(q)}$ through the module, and one backward pass for gradient computation. Assuming the more-computationally complex \ac{cm-ekf} is employed, the complexity order of these steps is of order $\mathcal{O}(KBP^2)$ each. However, as opposed to the usage of such methods for training an end-to-end \ac{dnn} as detailed in Subsection~\ref{subsec:bayes}, here $P$ represents the parameters of a compact module, which can be relatively small (e.g., $P<1000$ in our numerical study), as opposed to the overall \ac{dnn}. 
For comparison, training DeepSIC with $Q$ iterations and $K$ users via end-to-end \ac{cm-ekf} would result in complexity order of $\mySet{O}(B Q^2 K^3 P^2)$, while module-wise learning\footnote{Here, the per-module observation dimension is $B$,
so that the local Jacobian and noise covariance satisfy
$\myMat{H}_t \in \mathbb{R}^{B \times P}$ and 
$\myMat{R}_t \in \mathbb{R}^{B \times B}$,
where $P$ is the number of trainable parameters in the corresponding module.} requires order of $\mySet{O}(B Q K P^2)$.

Moreover, it is emphasized that our pipelining of the modular architectures results in the  {\em Predict-Update-Inference} steps in  Algorithm~\ref{alg:modular_update} being independent with each other, and can thus be computed in parallel. Accordingly, given simple parallel computation capabilities, the overhead in applying Algorithm~\ref{alg:modular_update} using \ac{cm-ekf} can be made to be of an order of $\mySet{O}(B P^2)$, where $P$ here is the number of module parameters (i.e., $\times Q K$ reduction due to parallelization).  When also combining \ac{dlr} approximations, its complexity reduces even further to a linear order of $\mySet{O}(B P)$.



\subsection{Discussion}
\label{subsec:discuss}
 
{The proposed modular online Bayesian learning framework addresses the three main challenges introduced in Subsection~\ref{subsec:problem}, namely, non-stationarity, latency/complexity, and nonlinearity (Challenges~\ref{itm:nonstat}--\ref{itm:nonlin}). By introducing modeling assumptions in a gradual and principled manner, our design enables efficient and rapid single-step online learning. Specifically, the modular architecture decomposes the detection task into compact and interpretable sub-networks, reducing the scale of adaptation. The Bayesian \ac{ssm} in parameter space captures smooth temporal dynamics of the receiver weights, naturally accommodating non-stationarity, and supports treating received pilots as streaming data. Posterior modeling with recursive Bayesian filtering allows each update to be carried out with minimal overhead, offering computationally tractable real-time adaptation in dynamic environments.}
\textcolor{NewColor2}{Accordingly, the Gaussian evolution model in parameter space and the use of local linearizations are modeling choices adopted to enable tractable and low-latency Bayesian filtering. While the reliability of these assumptions depends on the nature of channel variations, and may be less faithful in  abrupt environmental changes, such as sudden signal blockages or handovers, it is numerically shown in Section~\ref{sec:exp} that the resulting methodology still leads to rapid and efficient online adaptation.}

\textcolor{NewColor}{Modeling the receiver as a Bayesian neural network allows us to associate a distribution with the network parameters, whose covariance naturally quantifies the uncertainty in the current parameter estimate. Quantifying parameter space uncertainty enables the principled combination of new pilot observations with the previously learned model through recursive Bayesian filtering~\cite{blundell2015weight}. \emph{Parameter uncertainty} can be converted into \emph{predictive uncertainty} at the output of the receiver via ensembling of models sampled from the parameter space distribution~\cite{simeone2022} or via distribution methods~\cite{gawlikowski2023survey}. Therefore, the parameter uncertainty maintained by the Bayesian update rules serves exclusively as a modeling device that enables efficient online learning, supporting Bayesian filtering in dynamic communication environments.}

At the core of our methodology lies the departure from conventional \ac{sgd}-based training, which dominates deep learning but requires multiple iterations. By casting online learning as a tracking problem, we enable the use of recursive Bayesian filters as training algorithms. While nonlinear Kalman filtering methods have been considered for adapting small neural networks over 25 years ago~\cite{wan2000unscented}, their usage for learning was largely abandoned as modern machine learning shifted towards extremely large-scale models, where adaptation was only feasible through stochastic gradient based optimization. Recent attempts to reintroduce restricted Kalman-style learning into deep models, e.g.,~\cite{xia2025koala}, have primarily targeted offline settings. In contrast, our approach fuses Bayesian filtering with modular architectures, where each module can be individually evaluated and updated, thereby making single-step Bayesian learning not only possible but also computationally efficient.

\textcolor{NewColor}{Our framework gives rise to several extensions for future research. One direction is to explore additional Bayesian online learning formulations that explicitly assess uncertainty through more advanced metrics, such as coverage-based measures~\cite{zhang2024langevinized}, which could provide further insight into the reliability of learned parameter distributions in dynamic environments. 
A related direction is the investigation of ensemble-based Bayesian tracking methods, such as the ensemble Kalman filter and its stochastic variants~\cite{Evensen2003Ensemble,zhang2024langevinized}. While these approaches can provide improved posterior approximations and enhanced uncertainty quantification, their direct application to deep receivers is challenged by the need to maintain and process multiple model instances, leading to increased latency and computational overhead. Developing lightweight ensemble or hybrid filtering schemes that retain the benefits of sampling-based uncertainty representations while meeting the stringent real-time constraints of communication systems is therefore an interesting avenue for future work.
Another promising direction is the reduction of the dimensionality of the tracked parameter space by learning lower-dimensional latent representations of the receiver parameters, thereby enabling Bayesian tracking in a compact latent space~\cite{duran2024unifying}. 
Finally, one may consider alternative Bayesian tracking formulations that relax the Gaussian assumptions adopted in this work and allow for non-Gaussian parameter dynamics and unknown \ac{ssm} parameters, drawing inspiration from recent advances in Bayesian inference for stochastic dynamic systems~\cite{zhang2024extended,dong2023stochastic}.}


\section{Experimental Study}
\label{sec:exp}

This section numerically evaluates the online learning methodologies\footnote{Our code and all hyper-parameters used in this study can be found online at~\url{https://github.com/gusakovy/adaptive-deep-receivers}.} proposed in Section~\ref{sec:method}. Based on the common  experimental setup  detailed in Subsection~\ref{subsec:setup}, we evaluate four aspects of our methodology: 
$(i)$ Before assessing communication performance, we commence in Subsection~\ref{subsec:latency} with comparing the average online learning time of the different algorithms, to understand the real-time feasibility of each method; 
$(ii)$ In Subsection~\ref{subsec:lin_rot} we use a synthetic single user channel with a constant achievable rate to demonstrate the ability of single-step Bayesian adaptation to rapidly learn a near-optimal parameterization for a neural receiver, without any domain knowledge, and steadily track as the channel distribution changes in time; 
$(iii)$ In Subsection~\ref{subsec:modular_is_better} we showcase the advantages of modular architectures for online training on multi-user linear \ac{mimo} channels; 
$(iv)$ Finally, in Subsection~\ref{subsec:streaming_algo} we compare the performance of different algorithms that learn from streaming data, on both linear and nonlinear multi-user \ac{mimo} scenarios.

\subsection{Setup}
\label{subsec:setup}

\subsubsection{Receiver Architectures}
We utilize three different deep receivers. 
Our main \ac{dnn}-aided modular receiver architecture is based on DeepSIC~\cite{shlezinger2019deepsic}, with each module comprised of a compact two-layer \ac{mlp}. 
For a non-modular \ac{dnn}-based receiver we employ a ResNet-type architecture following \cite{honkala2021deeprx}.
For the single user setting (Subsection~\ref{subsec:lin_rot}), we use a fully-connected network with two layers.

\subsubsection{Learning Algorithms} 
\textcolor{NewColor}{Since the main objective of this work is to study rapid online learning for deep receivers, the numerical comparisons focus on different training algorithms applied to the same receiver architecture, focusing on  the impact of the learning methodology itself. Accordingly,}
we train the aforementioned architectures using the following  online learning algorithms:
\begin{itemize}
    \item {\bf{SGD}-$I\text{-}J$} - applies $I$ epochs of mini-batch \ac{sgd} with a batch size of $J$ on the cross entropy loss of a frequentist (non-Bayesian) model.
    
    \item {\bf{GD}-$I$} - applies $I$ iterations of \ac{gd} (one sample at a time)~\cite{Zinkevich2003Online}, starting from the previous $\dnnParam_{t-1}$,  using a frequentist model based on the cross-entropy loss.
    
    \item {\bf{BBB}-$I$} \acl{bbb}~\cite{blundell2015weight} adapted to online learning. For every sample, the update step applies $I$ \ac{gd} iterations on the online \ac{elbo} loss,
    \begin{align}
        \mathcal{L}_t(\bnnParam_t) =& 
        -\mathbb{E}_{\dnnParam_t \sim q_{\bnnParam_t}} \left[ \log p(\myVec{b}_t \mid \myVec{r}_t, \dnnParam_t) \right] \notag\\
        &+ D_\mathrm{KL}\left( q_{\bnnParam_t}(\dnnParam_t) \,\|\, q_{\bnnParam_{t | t\!-\!1}}(\dnnParam_t) \right),
        \label{eq:online_elbo}
    \end{align}
    i.e., for each iteration $i=1,\dots,I$, set
    \begin{equation}
        \bnnParam_{t}^{(i)} = \bnnParam_{t}^{(i-1)} - \alpha_t\nabla_{\bnnParam_{t}} \mathcal{L}_t\left(\bnnParam_{t}^{(i-1)}\right),
    \end{equation}
    where $\alpha_t$ is the learning rate. The iterations are initialized with $\bnnParam_t^{(0)}=\bnnParam_{t| t\!-\!1}$, and the expectation in (\ref{eq:online_elbo}) is approximated by linearization, which we found to be consistently more efficient and faster than the commonly used sample mean.

    \item{\bf{VD-EKF}/\bf{Lo-Fi}/\bf{CM-EKF}} - \ac{bong} implementation based on linearized Gaussian approximation (\ref{itm:LinGaus}). This leads to different \ac{ekf} variants, depending on whether \ac{dlr} (\ref{itm:DLR}) is assumed:  VD-EKF ($R=0$)~\cite{Chang2022on}; Lo-Fi ($0<R\ll P$)~\cite{Chang2023low}; and CM-EKF ($R=P$)~\cite{Tronarp2018CMEKF}.
    
    \item {\bf{BONG-EF}} - The \ac{bong} algorithm detailed in Subsection~\ref{subsec:bayes} using the \ac{ef} approach (\ref{itm:EF}). 
\end{itemize}
 
Among the compared algorithms, only \ac{sgd} does not use streaming data, meaning that it processes in batches and is allowed to look back at past samples to perform multiple epochs, while the other algorithms process one sample at a time.
It is also noted that the compared learning algorithms include methods that train frequentist \ac{dnn} parameters (\ac{sgd} and \ac{gd}) as well as ones that train the distribution parameters of \acp{bnn}. However, for fair comparison, once learning is concluded, performance is compared for all approaches using frequentist inference, which for the \acp{bnn} implies using the mean of the variational parameters as a plug-in approximation for inference.

\textcolor{NewColor2}{Unless stated otherwise, for all considered experimental scenarios, the state-space hyperparameters are set to $\gamma = 0.999$ and $\sigma^2 = 10^{-3}$. These values, which are shared across all evaluations, were based on experimental results from \cite{jones2024bayesian}, and further verified through empirical trials. In practice, $\sigma^2$ should be chosen to reflect the underlying channel dynamics, while the chosen value for $\gamma$ was found to be consistently appropriate, as it mainly ensures the variance of the stochastic process does not diverge.}


\subsection{Online Learning Latency}
\label{subsec:latency}
The main goal of our single-step online learning methodology is to enable adapting deep receivers to channel variations with an extremely minor overhead. 
We next show how each of the ingredients incorporated in our framework (single-step Bayesian learning, approximations \ref{itm:EF}-\ref{itm:DLR}, modular adaptation) translates into latency reduction. 
We consider a \ac{mimo} setting with $K=3$ users and $N=5$ antennas, using a modular (DeepSIC) and a non-modular (ResNet) deep receivers.
As our focus is on latency, we compare the learning algorithms that operate in a streaming manner (adapt on each incoming sample), 
under different forms of \ac{dlr} limitations $R$ (as \ref{itm:DLR} can be combined with any Bayesian adaptation method). All models were trained using JAX on the same platform, an Intel i7 12th Gen CPU, i.e., on a standard laptop without dedicated \ac{ai} accelerators. The resulting average latencies are reported in Table~\ref{tab:runtimes}. 

{Among the EKF-based methods (\ref{itm:LinGaus}), \ac{vd-ekf}, which enforces a diagonal covariance, achieves the lowest latency. Lo-Fi, which uses a diagonal + rank-$R$ covariance structure, introduces a moderate overhead compared to \ac{vd-ekf} but remains in the same latency regime as GD-$10$ while being considerably more expressive. The full covariance variant \ac{cm-ekf} incurs a considerable increase in latency.
The \ac{ef} approximation (\ref{itm:EF}) reduces the latencies of the \ac{bong} algorithm for all covariance types, at the cost of stability and performance reduction, as shown in the following subsections.}

{The effect of modularity is also evident: DeepSIC consistently achieves sub-millisecond adaptation, whereas the same update rules applied to a dense ResNet quickly become impractical. This highlights that combining modular architectures with single-step Bayesian adaptation is essential to meet real-time constraints.
\textcolor{NewColor2}{Beyond execution time, an important measure is the memory footprint of the proposed Bayesian adaptation. The primary storage requirement for full-rank methods stems from the covariance matrix associated with each module’s parameters, which scales as $\mathcal{O}(P^2)$, (recall that $P$ denotes the number of trainable parameters per module). By utilizing a modular architecture where each detection unit is a relatively small neural network (e.g., with around $P \approx 500$ parameters), the memory requirement is kept to approximately 1 MB per module. This modularity is crucial for scalability: in a non-modular, monolithic receiver architecture, the total number of parameters would be significantly larger, forcing the $\mathcal{O}(P^2)$ storage of the covariance matrix into the gigabyte range, resulting in out-of-memory errors noted in Table~\ref{tab:runtimes}.}}

\begin{table}[htbp]
    \centering
    \fontsize{7pt}{9pt}\selectfont
    \begin{tabular}{|c|c|cc|}
        \hline
        \rowcolor[HTML]{C0C0C0}
        \cellcolor[HTML]{C0C0C0} &
        \cellcolor[HTML]{C0C0C0} &
        \multicolumn{2}{c|}{\cellcolor[HTML]{C0C0C0}\textbf{Latency/sample} [{\em mSec}]} \\ \cline{3-4}
        
        \rowcolor[HTML]{C0C0C0}
        \multirow{-2}{*}{\cellcolor[HTML]{C0C0C0}\textbf{Algorithm}} &
        \multirow{-2}{*}{\cellcolor[HTML]{C0C0C0}\textbf{\begin{tabular}[c]{@{}c@{}}$\myMat{\Sigma}_t$ DLR \\ rank $R$ (\ref{itm:DLR})\end{tabular}}} &
        \multicolumn{1}{c|}{\cellcolor[HTML]{C0C0C0}\textbf{DeepSIC}} &
        \textbf{ResNet} \\ \hline

        \begin{tabular}[c]{@{}c@{}}GD-$I$ \\ $I=10$\end{tabular} & 
        N/A &
        \multicolumn{1}{c|}{0.268} &
        0.528 \\ \hline

        & Full: $R=P$ &
        \multicolumn{1}{c|}{OOM} &
        OOM \\ \cline{2-4}

        \multirow{-2}{*}{\begin{tabular}[c]{@{}c@{}}BBB-$I$ \\ $I=10$\end{tabular}} &
        Diag: $R=0$ &
        \multicolumn{1}{c|}{0.644} &
        14.12 \\ \hline

        & Full: $R=P$ (CM-EKF) &
        \multicolumn{1}{c|}{2.942} &
        OOM \\ \cline{2-4}

        & DLR: $R=10$ (Lo-Fi) &
        \multicolumn{1}{c|}{0.356} &
        8.117 \\ \cline{2-4}

        \multirow{-3}{*}{\begin{tabular}[c]{@{}c@{}}BONG + \ref{itm:LinGaus} \\ (\ac{ekf}-type)\end{tabular}} &
        Diag: $R=0$ (VD-EKF) &
        \multicolumn{1}{c|}{0.103} &
        4.526 \\ \hline

        & Full: $R=P$ &
        \multicolumn{1}{c|}{1.712} &
        OOM \\ \cline{2-4}

        & DLR: $R=10$ &
        \multicolumn{1}{c|}{0.319} &
        1.443 \\ \cline{2-4}

        \multirow{-3}{*}{\begin{tabular}[c]{@{}c@{}}BONG + \ref{itm:EF} \\ (\ac{ef})\end{tabular}} &
        Diag: $R=0$ &
        \multicolumn{1}{c|}{0.083} &
        1.230 \\ \hline
    \end{tabular}

    \vspace{0.2cm}
    \caption{Average training iteration time per sample, $K,N=(3,5)$; \ac{oom} denotes methods that could not be executed.}
    \label{tab:runtimes}
    \vspace{-0.5cm}
\end{table}


\subsection{Linear Rotation Channel}
\label{subsec:lin_rot}
Here, we consider a simple linear rotation channel with a single user and QPSK symbols. The channel for time $t$ is modeled as a noisy rotation by $\varphi_t $ radians, i.e., 
\begin{equation}
    \myVec{r}_{t} = \begin{bmatrix} \cos(\varphi_t) & -\sin(\varphi_t)\\ \sin(\varphi_t) & \cos(\varphi_t) \end{bmatrix} \myVec{s}_{t} + \myVec{u}_{t},
    \label{eqn:LinChannel}
\end{equation}
where $\myVec{u}_{t}$ is Gaussian white noise with variance $\sigma_u^2=1/16$. The rotation angle increases linearly with parameter $\alpha= 2.5\cdot 10^{-4}$, i.e., $\varphi_t = 2\pi\alpha t$. 

The purpose of this setting is to evaluate our online learning mechanism in a simple setup where model-based receiver algorithms can be applied and approach optimality. Specifically, for the channel in \eqref{eqn:LinChannel}, the error rate of the optimal decoder is constant, and can be derived using the \ac{map} rule. While the \ac{map} rule requires knowledge of the instantaneous channel, it can be efficiently approached in this setting using  \ac{nlms} channel estimation followed by equalization and minimum distance decoding.
The neural receivers are feed-forward networks with a single hidden layer of dimension 10 followed by a ReLU activation function, and an output layer followed by an element-wise Sigmoid operator that maps to soft bit-level estimates. We compare here  \ac{cm-ekf}-based learning with standard \ac{sgd}. For both \ac{nlms} and \ac{sgd}, we manually optimized the learning rates.

\begin{figure}
    \centering
    \includegraphics[width=0.5\textwidth]{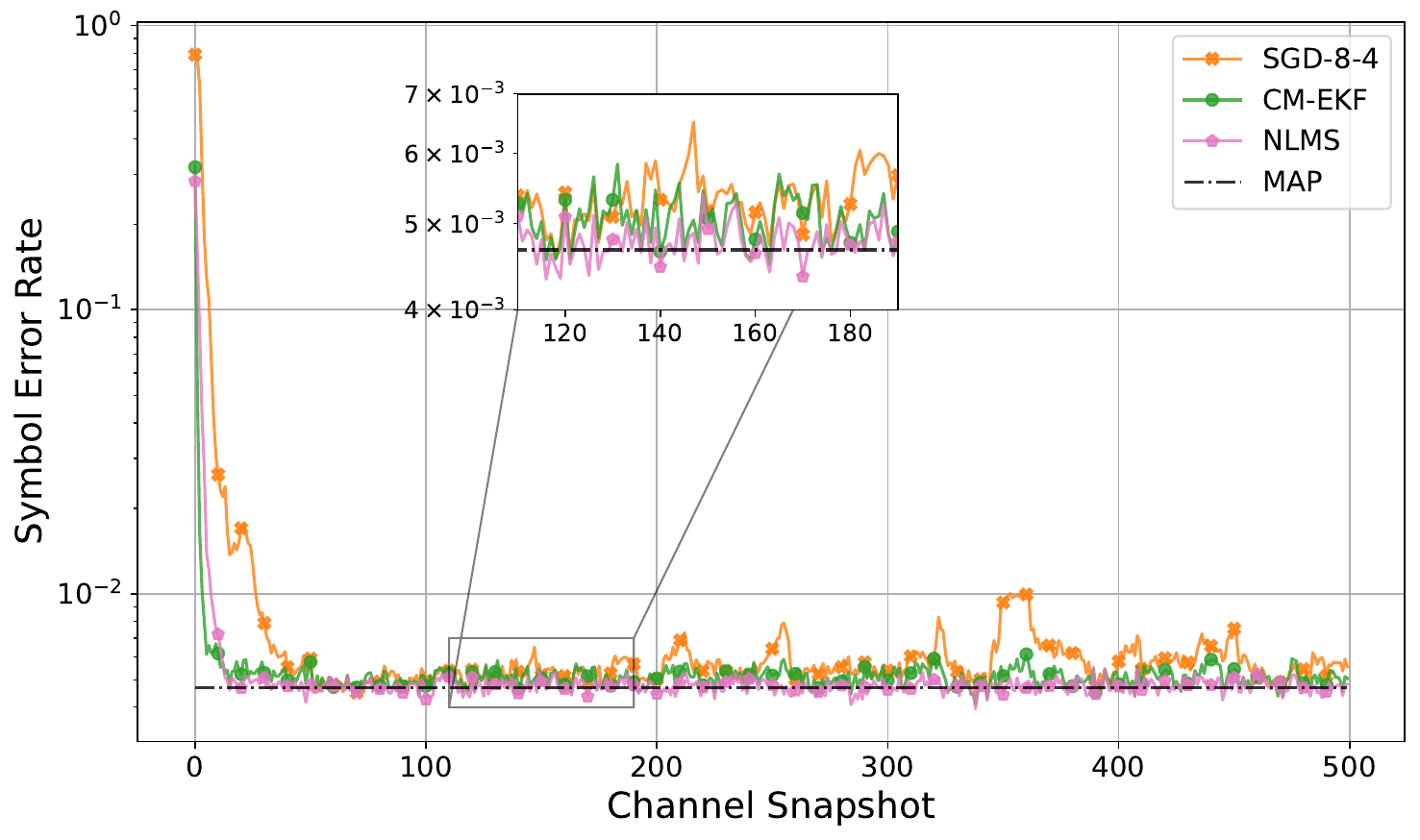}
    \vspace{-0.2cm}
    \caption{Symbol error rates, linear rotation channel.}
    \label{fig:lin_rot_ser}
\end{figure}

\begin{figure*}
    \centering
    \centerline{\includegraphics[width=17cm]{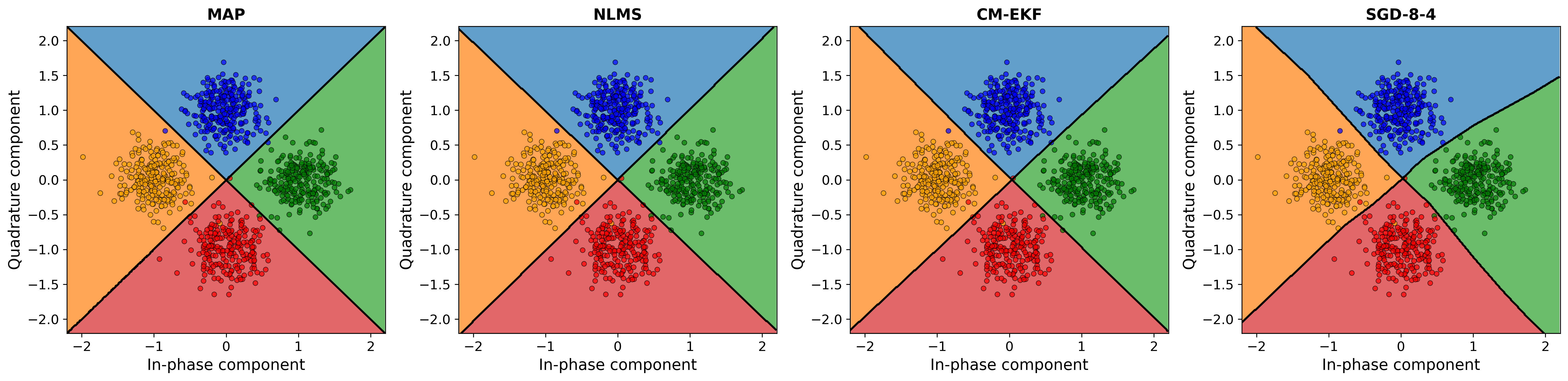}}
    \caption{Decision zones after 500 blocks (45$^\circ$ counter-clockwise rotation) for four decoders - optimal decoding using the \ac{map} rule, channel estimation using \ac{nlms} followed by maximum likelihood decoding, and two neural decoders trained using \ac{cm-ekf} and \ac{sgd}-8-4.}
    \label{fig:decision_zones}
\end{figure*}

Fig.~\ref{fig:lin_rot_ser} illustrates the symbol error rates during $500$ channel snapshots, using $16$ pilots per snapshot. The average error rate of \ac{cm-ekf} is lower than that of learning rate optimized \ac{sgd} performing $8$ epochs with batch size $4$. Furthermore, it converges faster than both \ac{sgd}-8-4 and \ac{nlms}, taking only $6$ snapshots to reach within $0.2\%$ of the optimal symbol error rate, compared to around $12$ it takes \ac{nlms} and $40$ it takes \ac{sgd}. Training using \ac{sgd} is also less stable, with a significantly higher peak error when tracking.
Fig.~\ref{fig:decision_zones} visualizes the decision zones of each receiver after $500$ blocks. The \ac{cm-ekf} trained decoder closely matches the \ac{map} decision zones, underlining the ability of the Bayesian receiver to generalize well, while the \ac{sgd} trained decoder noticeably distorts the decision zones, overfitting to noisy observations.

\subsection{Modular vs. Non-Modular Architectures}
\label{subsec:modular_is_better}

\begin{figure}
    \centering
    \includegraphics[width=0.47\textwidth]{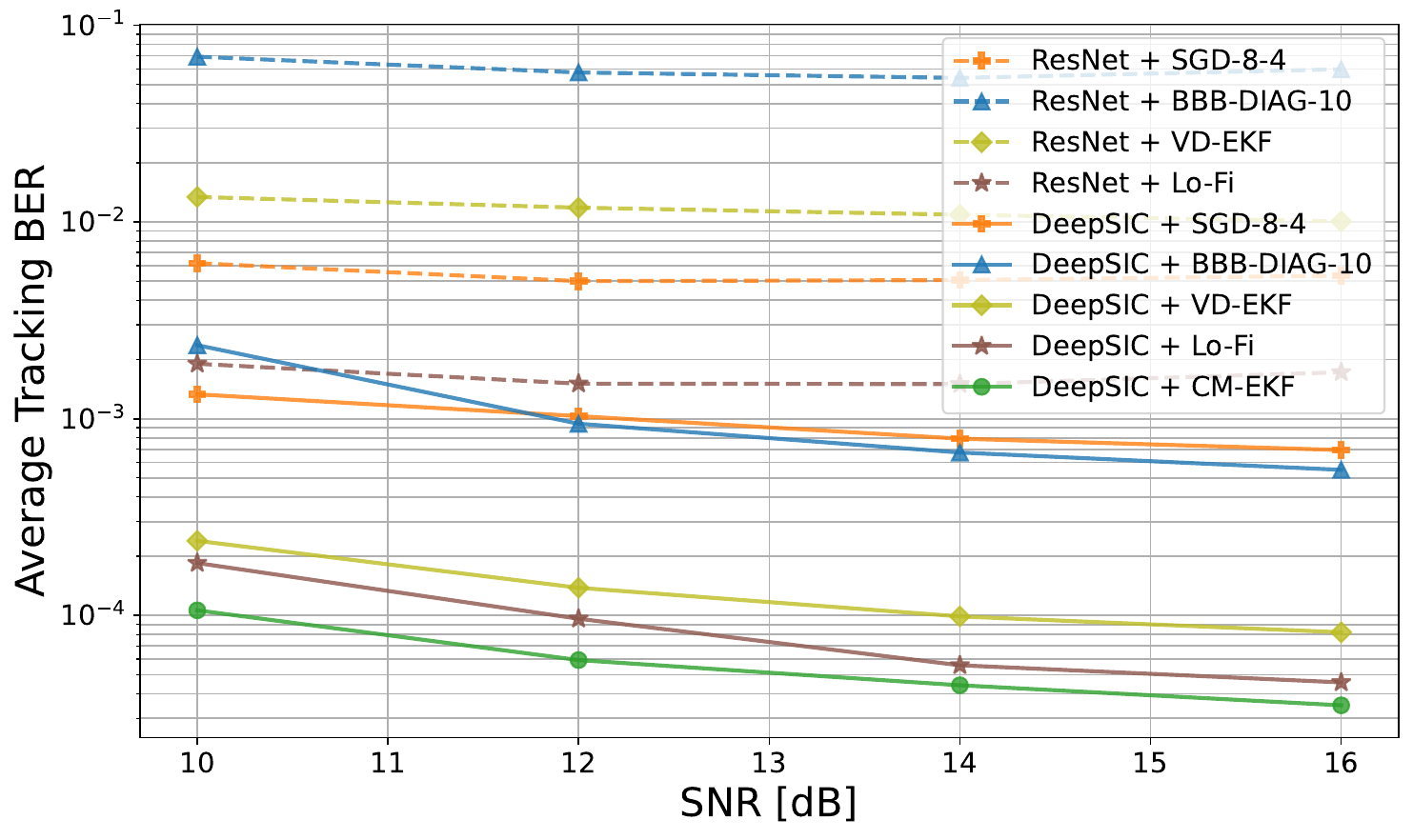}
    \caption{Online learning for modular vs. non-modular architectures.}
    \label{fig:mod_vs_nonmod}
\end{figure}

We proceed to demonstrate the usefulness of modular architectures for enabling efficient online learning using our Bayesian methodology. To that aim, we compare the modular DeepSIC architecture with a black-box \ac{dnn} based on a ResNet architecture, known to be suitable for deep receivers~\cite{honkala2021deeprx}. The \ac{dnn} is comprised of an input layer followed by ReLU activation, $4$ residual layers~\cite{He2016Resnet}, and an output layer followed by an elementwise Sigmoid operator. The input of this ResNet is the received signal $\myVec{r}_t$, and the output is a vector of soft bit-level estimates for all users, $\myVec{\ell}_t$.

The data for this experiment is generated from the QuaDRiGa \ac{mimo} geometry-based simulator~\cite{jaeckel2014quadriga}, using the 3GPP \acl{los} configuration, having $K=3$ single-antenna users in motion transmit to a $N=5$ antenna receiver. 
We take $100$ consecutive channel snapshots at constant intervals, and $64$ symbols are transmitted for each snapshot. The first $4$ snapshots are used for synchronization, i.e., $T_\text{sync}=256$, and for the rest transmit $16$ pilots per snapshot with the intention of providing sufficient training data. All results are averaged over 10 trials.

Fig.~\ref{fig:mod_vs_nonmod} illustrates the average \ac{ber} vs. \ac{snr} for modular and non-modular architectures. Both ResNet and DeepSIC were trained online using SGD-$8\text{-}4$ on pilots from each channel snapshot separately, and \acs{bbb}-DIAG-$10$, \ac{vd-ekf}, Lo-Fi with rank $R=10$, and \ac{cm-ekf} on streaming data. The ResNet architecture could not be trained using \ac{cm-ekf} due to the size of the network ($P=32,766$). The results highlight the benefits of modular architectures for online adaptation. For the non-modular ResNet receiver, \acs{bbb} fails to train the network entirely, resulting in consistently poor performance across all SNR values. Training with \ac{vd-ekf} and \ac{sgd}-$8$-$4$ allows the ResNet to synchronize, but it still struggles to maintain low error rates during the tracking period. Lo-Fi improves the performance in the tracking phase, but it still remains lower than all methods when applied to the modular DeepSIC receiver. In contrast, each module of the modular DeepSIC architecture is a much smaller network ($P=458$) tasked with decoding a single user, which makes synchronization easier and substantially improves online adaptability. The \ac{ekf} variants perform markedly better than 10 iterations of either \ac{sgd} and \ac{bbb}-DIAG. \ac{vd-ekf} is the least expressive of the \ac{ekf} variants but has lower runtime than \ac{bbb}-DIAG-$10$. Lo-Fi has a runtime comparable to that of \ac{bbb}-DIAG-$10$ with improved error rate, and \ac{cm-ekf} has the lowest error rates across all \ac{snr} values, but is slower than \ac{bbb}-$10$.

\subsection{Streaming-Data Algorithms}
\label{subsec:streaming_algo}

\begin{figure*}
    \centering
    \centerline{\includegraphics[width=17cm]{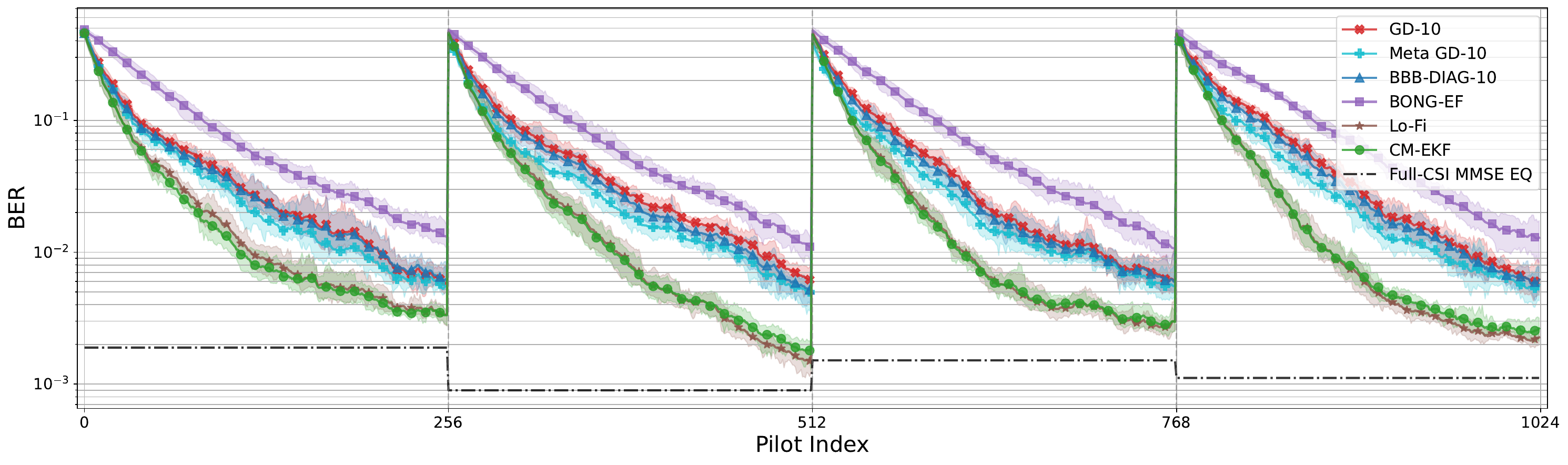}}
    \caption{BER vs. pilot index at SNR=$4$ dB. Synchronization from streaming data, with four different Sionna narrowband linear channels.}
    \label{fig:sionna_abrupt}
\end{figure*}

Having demonstrated the usefulness of combining modular architectures with online Bayesian learning, we proceed to compare algorithms that learn online from streaming data with DeepSIC. Specifically, we consider the Bayesian algorithms \acs{bbb}-DIAG-$10$, Lo-Fi with rank $R=10$, \ac{cm-ekf} and \ac{bong}-\ac{ef}, as well as the non-Bayesian \ac{gd}-$10$ algorithm. All results are averaged over 10 trials.

The data for this experiment is generated from three sources:
\begin{enumerate}
    \item[$(i)$] A batch of static narrowband channels generated using the Sionna~\cite{Sionna} differentiable link level simulator using the \acf{umi} channel model, having $K=4$ single antenna users, randomly scattered within a cell sector, transmit to a single $N=8$ antenna receiver. We take 4 different channels and use $T_\text{sync}=256$ pilots for synchronization to each of them.

    \item[$(ii)$] The COST2100~\cite{liu2012cost} geometry-based channel simulator, using the dynamic "indoor closely spaced users" configuration, having $K=3$ users transmit to a $N=5$ receiver antenna. We take $300$ consecutive channel snapshots at constant intervals, each comprised of $64$ symbols, with the first $2$ snapshots used for synchronization, i.e., $T_\text{sync}=128$, while the remaining snapshots include only $2$ pilots each;

    \item[$(iii)$] A nonlinear channel acquired by applying a nonlinear transformation, and particularly a $\tanh$ to a linear channel generated from the QuaDRiGa simulator using a configuration similar to the one used in Subsection~\ref{subsec:modular_is_better}. 
    This operation may represent distortions induced by the receiver acquisition hardware. 
    We take $100$ channel snapshots of $64$ symbols, with the first $4$ used for synchronization, i.e., $T_\text{sync}=256$, while the remaining snapshots include only $16$ pilots.
\end{enumerate}
  
{For the Sionna channel synchronization experiment (Fig.~\ref{fig:sionna_abrupt}), we look at synchronization performance for 4 different channels with 256 pilots transmitted for each of them. Since within every window of 256 pilots the channel is constant, we report the \ac{ber} of the \ac{mmse} decoder with full \ac{csi} for reference. In addition, we also use \ac{maml}~\cite{Finn2017Maml} to learn an initialization for quick adaptation to new channels. To do this, we generate additional channels from the same model, and use meta learning to train an initial parameterization for the DeepSIC receiver, which minimizes the \ac{ber} during synchronization with \ac{gd}-10. The learned initialization reduces the \ac{ber} of \ac{gd}-10 throughout the entire synchronization period, helping it synchronize faster than \ac{gd}-10 with random initialization, and \ac{bbb}-DIAG-10. Even so, the \ac{ekf}-based methods Lo-Fi and \ac{cm-ekf}  synchronize much faster than the iterative methods, both achieving near optimal error rates using only $256$ samples. \ac{bong}-\ac{ef}, on the other hand, synchronizes slower than all other methods, likely attributed to the noisy \ac{ef} approximation.}

\begin{figure*}
    \begin{minipage}[b]{0.49\linewidth}
      \centering
      \centerline{\includegraphics[width=7.5cm]{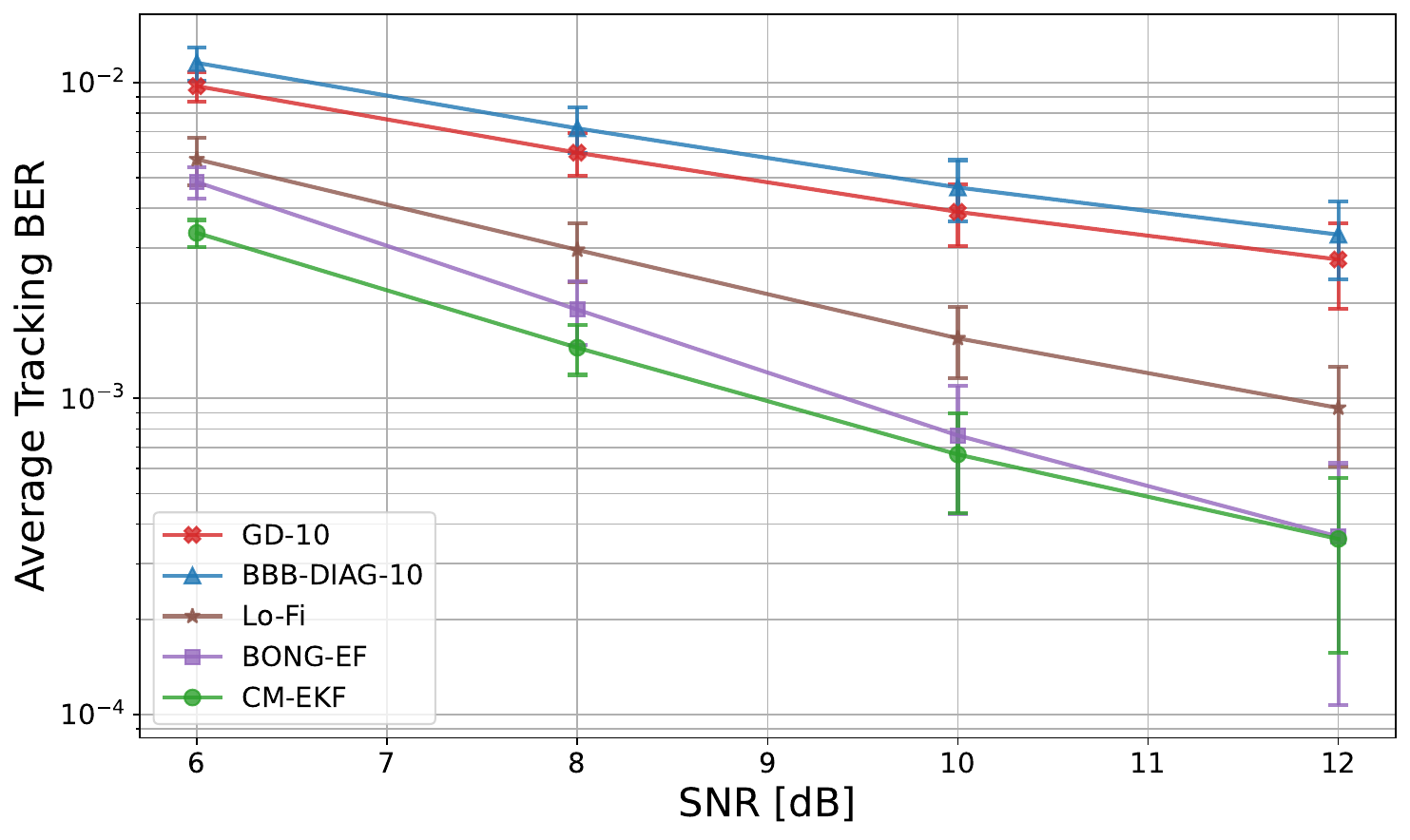}}
    \end{minipage}
    \begin{minipage}[b]{0.49\linewidth}
      \centering
      \centerline{\includegraphics[width=7.5cm]{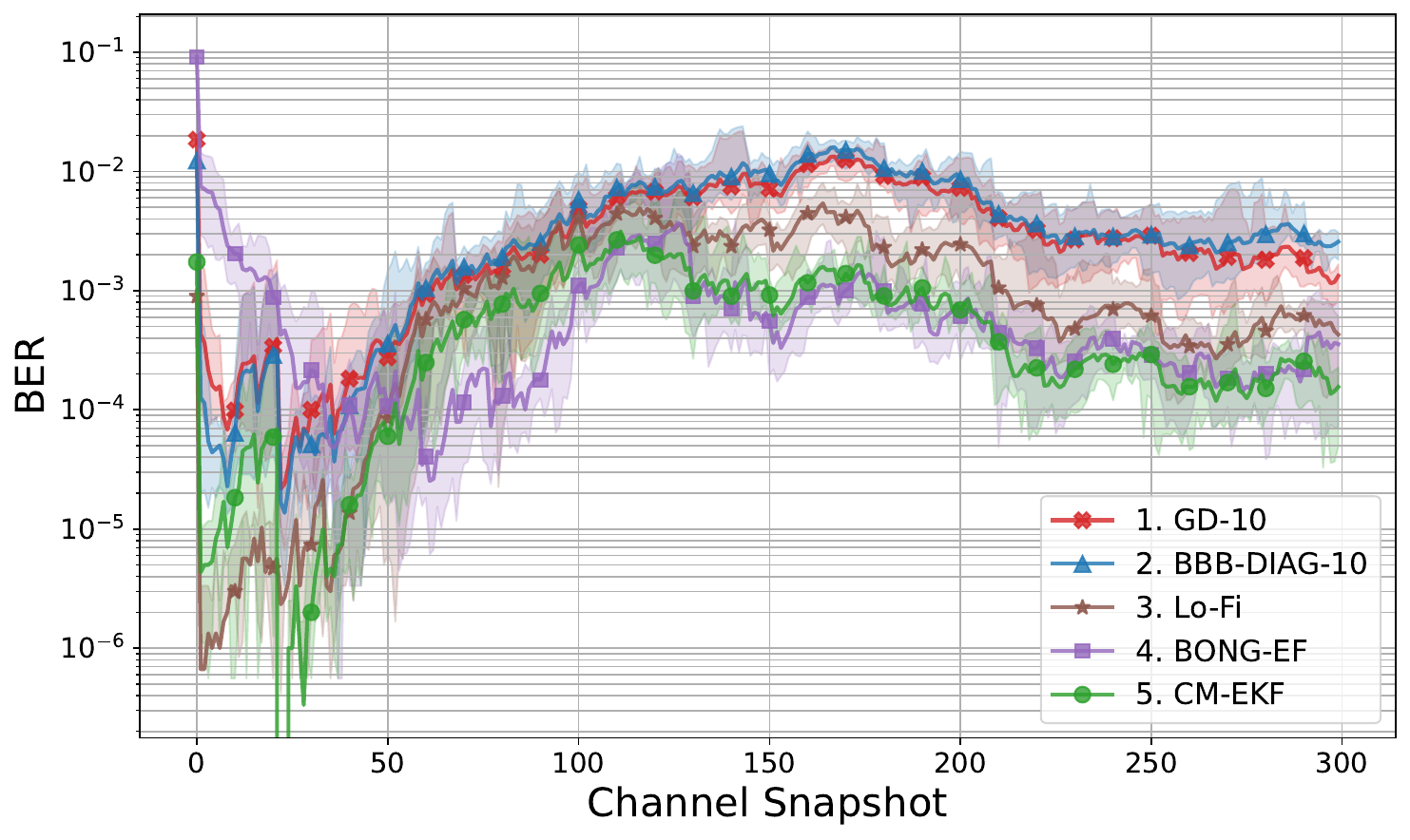}}
    \end{minipage}
    \caption{Learning from streaming data, COST2100 channel. Left: Average \ac{ber} vs. SNR; Right: \ac{ber} vs channel snapshot at SNR$=10$ dB.}
    \label{fig:cost2100_linear}
\end{figure*}

\begin{figure*}
    \begin{minipage}[b]{0.49\linewidth}
      \centering
      \centerline{\includegraphics[width=7.5cm]{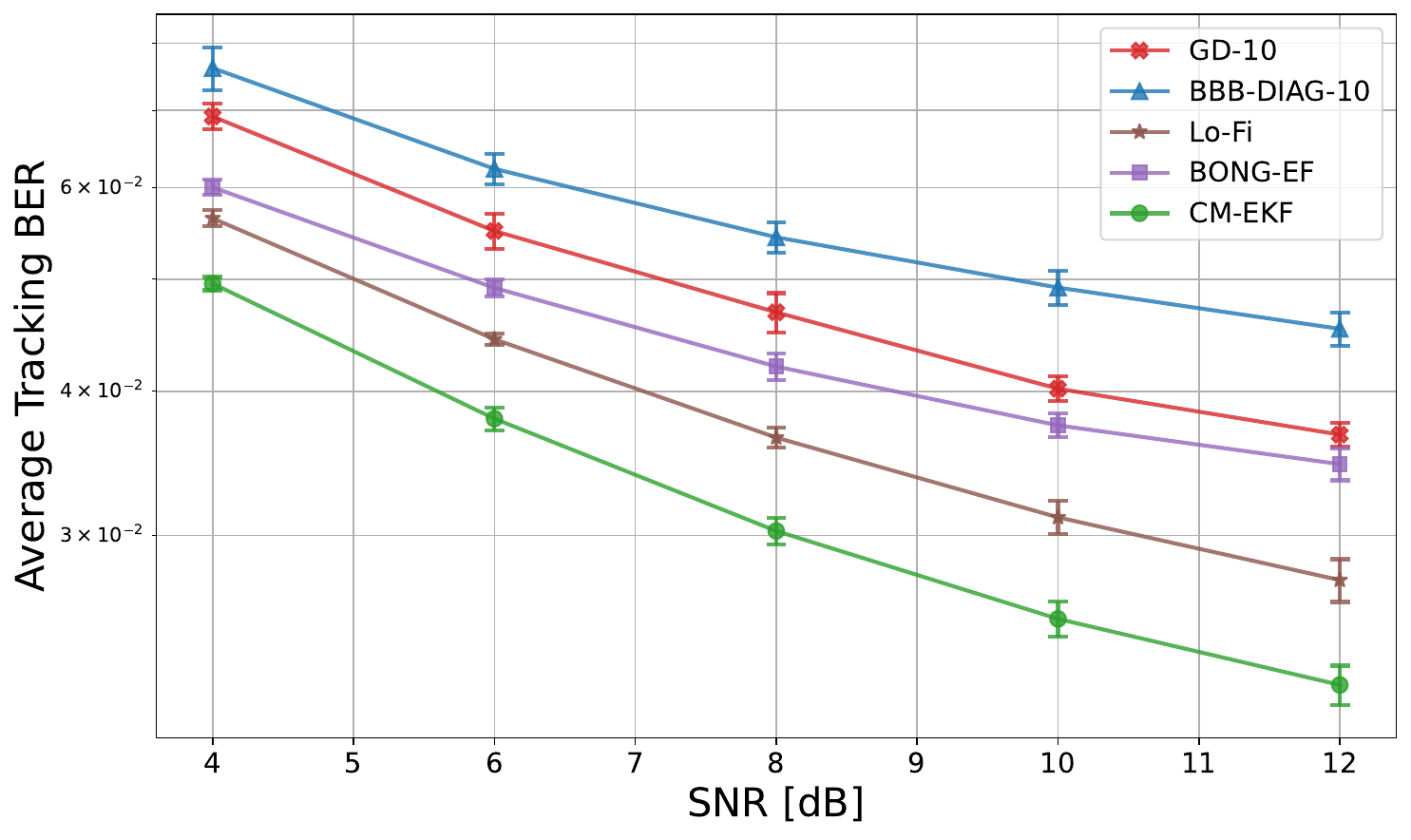}}
    \end{minipage}
    \begin{minipage}[b]{0.49\linewidth}
      \centering
      \centerline{\includegraphics[width=7.5cm]{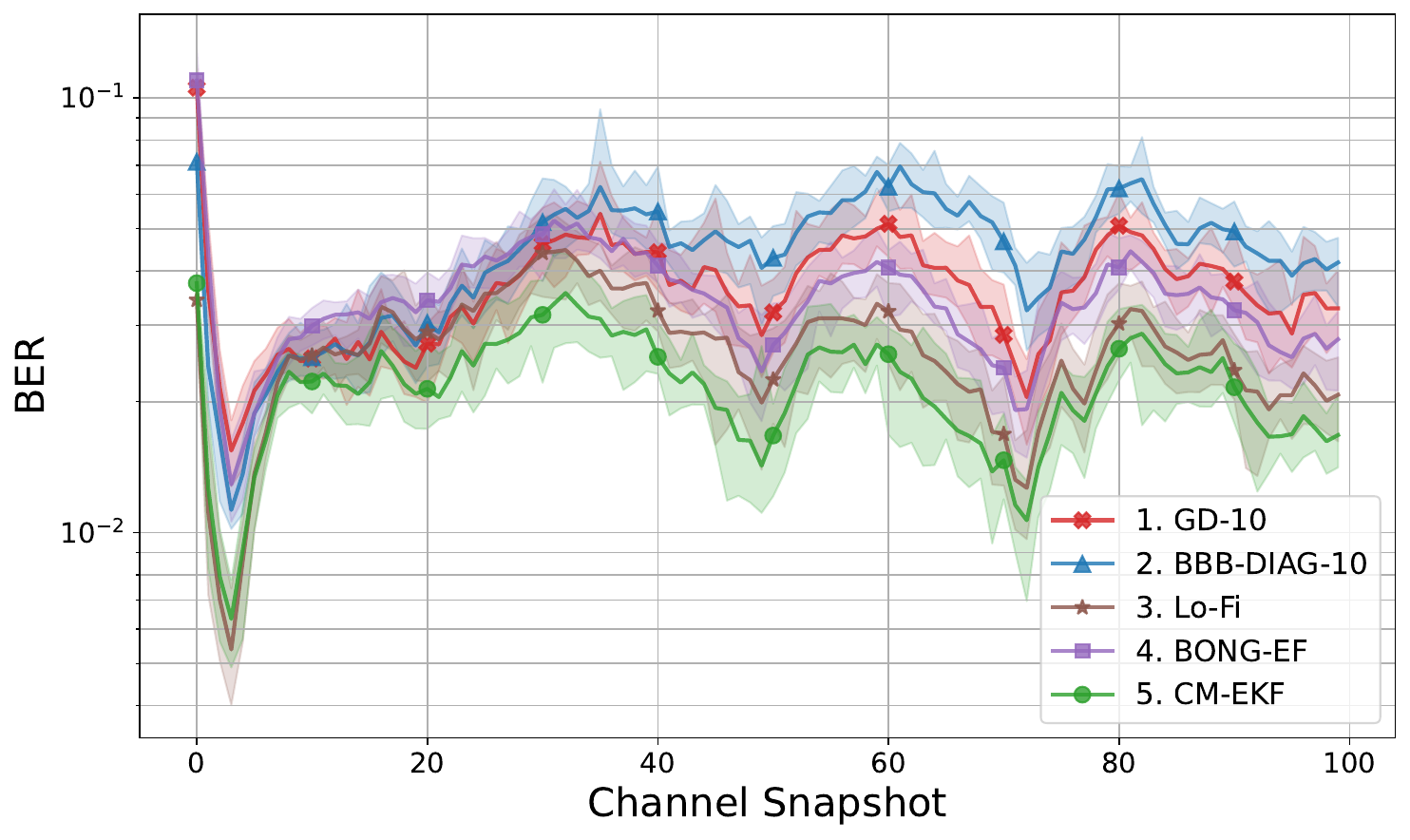}}
    \end{minipage}
    \caption{Learning from streaming data,  nonlinear QuaDRiGa channel. Left: Average \ac{ber} vs. SNR; Right: \ac{ber} vs channel snapshot at SNR$=12$ dB.}
    \label{fig:quadriga_nonlinear}
\end{figure*}

For the COST2100 channel experiment (Fig.~\ref{fig:cost2100_linear}), the \ac{cm-ekf} outperforms all other methods across all SNR values, with \ac{bong}-\ac{ef} coming in as a close second at high \ac{snr}. These full-covariance single-step approaches are consistently able to leverage the limited observations to more effectively adapt the receiver parameters to the varying channel conditions. However, \ac{bong}-\ac{ef} requires significantly more time to synchronize with the channel, due to the inherent instability introduced by using the \ac{ef} approximation instead of the Hessian. By contrast, Lo-Fi synchronizes faster than any other method, thanks to its lighter yet still expressive parameterization. When tracking, Lo-Fi does fall short of \ac{cm-ekf} and \ac{bong}-\ac{ef}, but runs a lot faster. \acs{bbb}-DIAG-$10$ and \ac{gd}-$10$ show comparable performance: both manage to adapt during the synchronization period, but fail to sustain low error rates in the subsequent tracking phase. This degradation can be attributed to the noisy parameter updates that occur when operating with scarce data.

\begin{figure*}
    \begin{minipage}[b]{0.49\linewidth}
      \centering
      \centerline{\includegraphics[width=7.5cm]{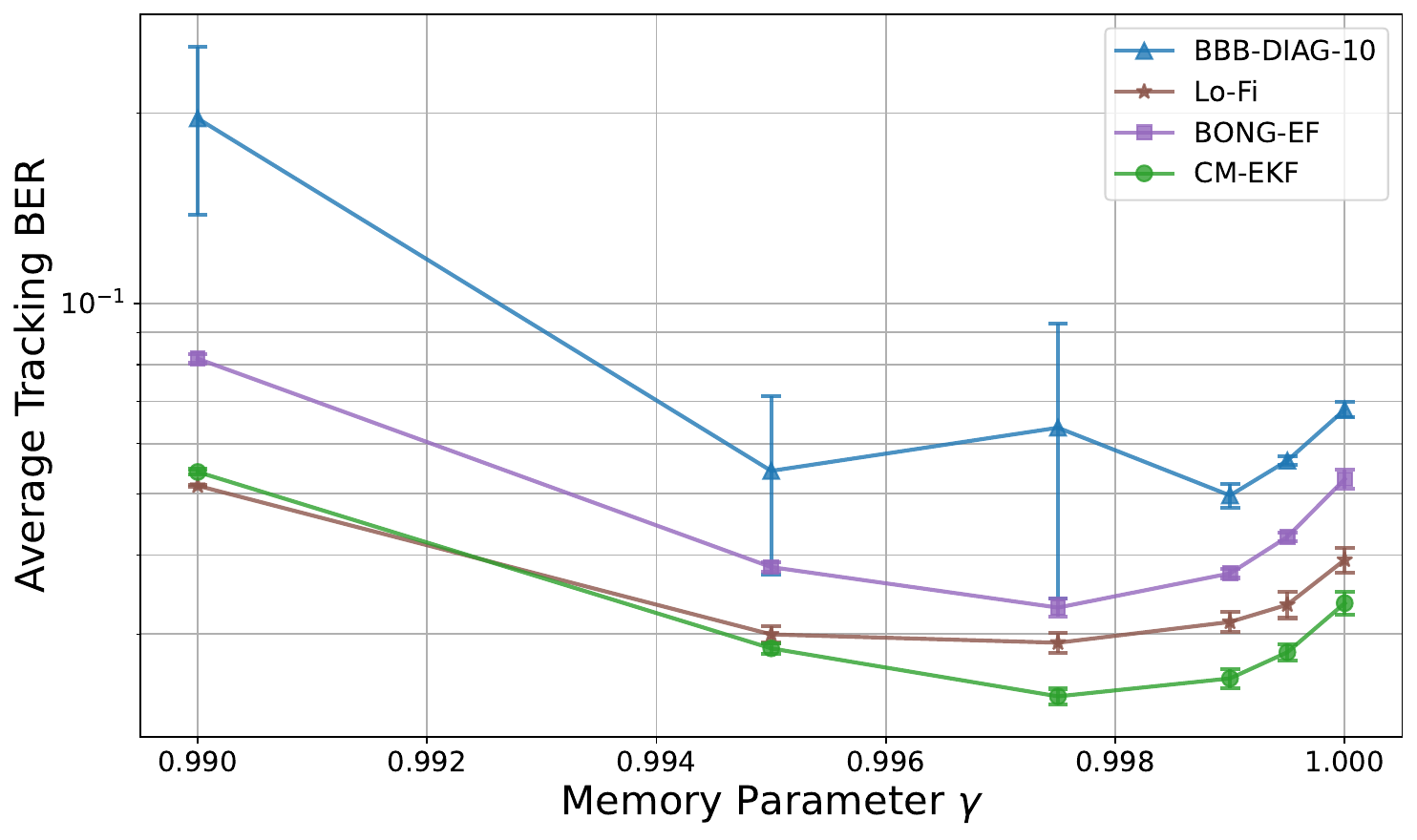}}
    \end{minipage}
    \begin{minipage}[b]{0.49\linewidth}
      \centering
      \centerline{\includegraphics[width=7.5cm]{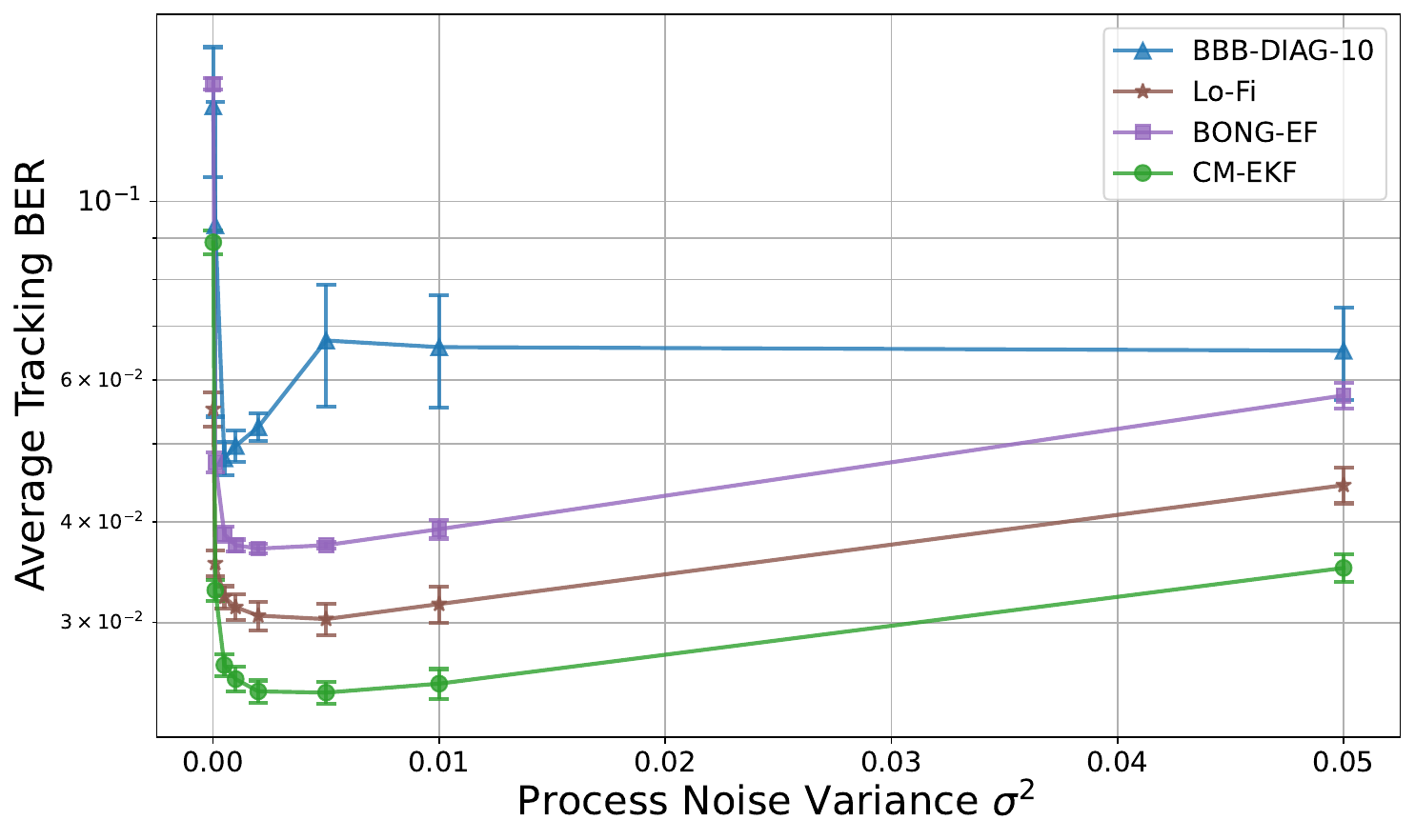}}
    \end{minipage}
    \caption{Sensitivity to state evolution parameters, nonlinear QuaDRiGa channel at SNR$=12$ dB. Left: Average \ac{ber} vs. $\gamma$; Right:  Average \ac{ber} vs. $\sigma^2$.}
    \label{fig:quadriga_sensitivity}
\end{figure*}

In the nonlinear scenario (Fig.~\ref{fig:quadriga_nonlinear}), the channel conditions are substantially harsher, demanding highly accurate parameter updates. In this setting, \ac{cm-ekf} decisively outperforms all other methods across all SNR values by several standard deviations. It not only synchronizes more rapidly than most of the alternatives except Lo-Fi, but it also consistently maintains superior performance throughout the entire tracking period. In comparison, Lo-Fi exhibits the second-best error rates during tracking across all \ac{snr} values, further underlining the trade-off between complexity and tracking performance. \ac{bong}-\ac{ef} exhibits slower synchronization and only marginally surpasses the performance of \ac{gd}-$10$. Meanwhile, \acs{bbb}-DIAG-$10$ struggles to deliver stable tracking, failing to sustain reliable performance under the severe nonlinear conditions.

\textcolor{NewColor}{We conclude our numerical study by examining the sensitivity of the proposed online learning framework to the state evolution hyperparameters. To that aim, we conduct an additional experiment using the nonlinear QuaDRiGa scenario at \ac{snr}$=12$ dB, and evaluate the average \ac{ber} achieved by all the streaming-data algorithms while varying the memory parameter $\gamma$ and the process noise variance $\sigma^2$ governing the state evolution model. The results, shown in Fig.~\ref{fig:quadriga_sensitivity}, indicate that similar ranges of hyperparameters $\gamma$ and $\sigma^2$ provide favorable performance across all the considered algorithms. This suggests that these parameters primarily reflect the dynamics of the underlying channel evolution, rather than the specific choice of the online learning algorithm.}

\textcolor{NewColor}{Overall, the results highlight the interplay between the different complexity–performance mechanisms discussed in Subsection~\ref{subsec:bayes}. 
First, the structure imposed on the covariance matrix determines the balance between expressiveness and computational cost. Full-covariance tracking with \ac{cm-ekf} consistently yields the most reliable adaptation, particularly in challenging or nonlinear scenarios, while the \ac{dlr} and diagonal approximations (\ref{itm:DLR}) provide a favorable compromise between latency and tracking accuracy, often synchronizing the fastest; 
Second, replacing the Hessian with the empirical Fisher approximation (\ref{itm:EF}) further reduces computational overhead, but may introduce noisy updates that slow synchronization and degrade stability compared to using \ac{ekf}-type methods (\ref{itm:LinGaus}). 
Accordingly, practitioners seeking the highest robustness should favor EKF-based variants (with \ac{cm-ekf} offering the best performance and Lo-Fi providing a fast and stable alternative), whereas the empirical Fisher variant is most appropriate in settings where minimizing online learning latency is the primary concern and some loss in stability can be tolerated.}


\section{Conclusions}
\label{sec:conclusions}

In this work, we proposed a modular online Bayesian learning framework for deep receivers operating over time-varying channels. By casting online learning as a state-space tracking problem and adopting recursive Bayesian updates, we demonstrated that reliable one-shot adaptation can be achieved, replacing multi-epoch training with single-step updates. Furthermore, by leveraging modular receiver architectures such as DeepSIC, we showed how module-wise Bayesian adaptation from streaming pilots enables parallelization and pipelining, thereby substantially reducing both complexity and latency. 


\bibliographystyle{IEEEtran}
\bibliography{IEEEabrv, ref}

%
\begin{IEEEbiography}[{\includegraphics[width=1in,height=1.25in,clip,keepaspectratio]{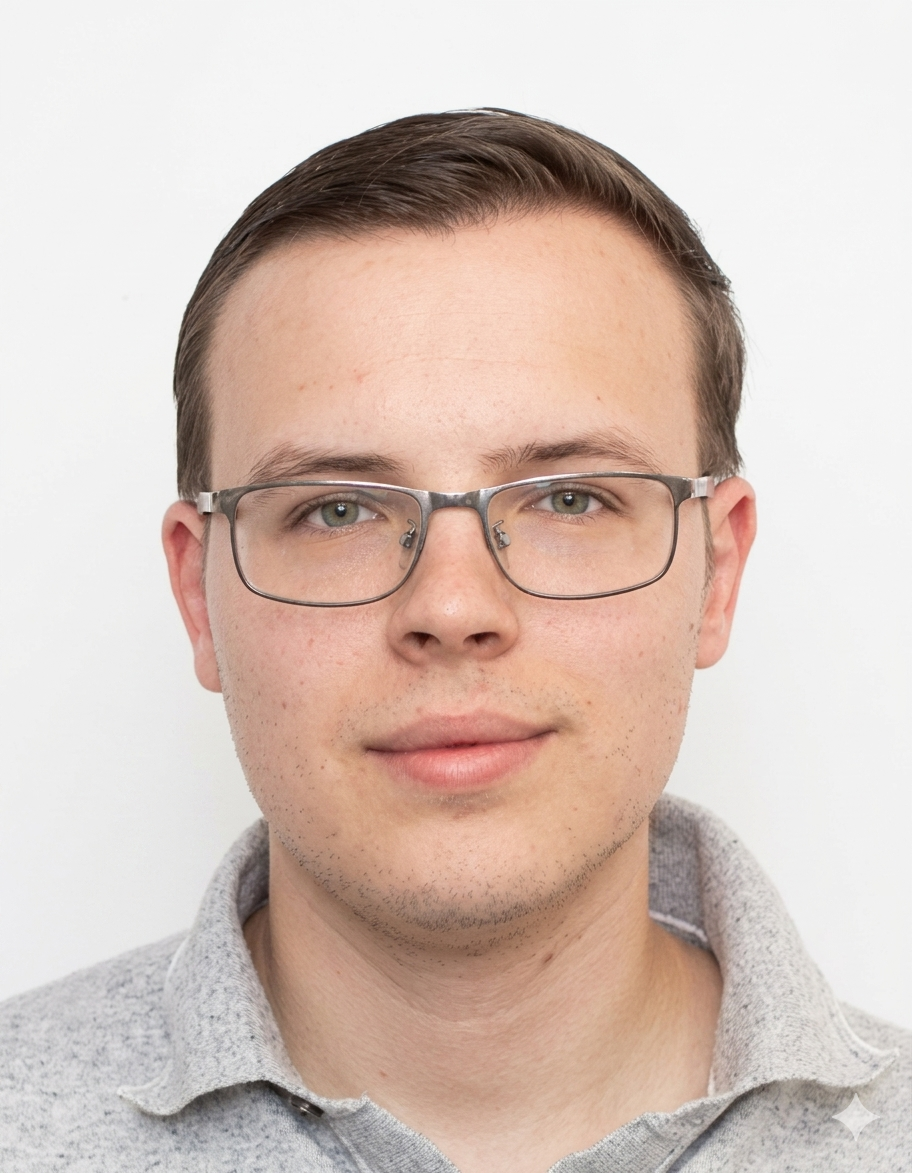}}]{Yakov Gusakov}
received his dual B.Sc. in Electrical Engineering and Mathematics (with highest honors) in 2024, and his M.Sc. in Electrical Engineering in 2025, both from Ben-Gurion University of the Negev, Be’er Sheva, Israel. His M.Sc. research, conducted under the supervision of Prof. Tirza Routtenberg and Dr. Nir Shlezinger, focuses on Bayesian online learning for wireless deep receiver design.
\end{IEEEbiography}

\begin{IEEEbiography}[{\includegraphics[width=1in,height=1.25in,clip,keepaspectratio]{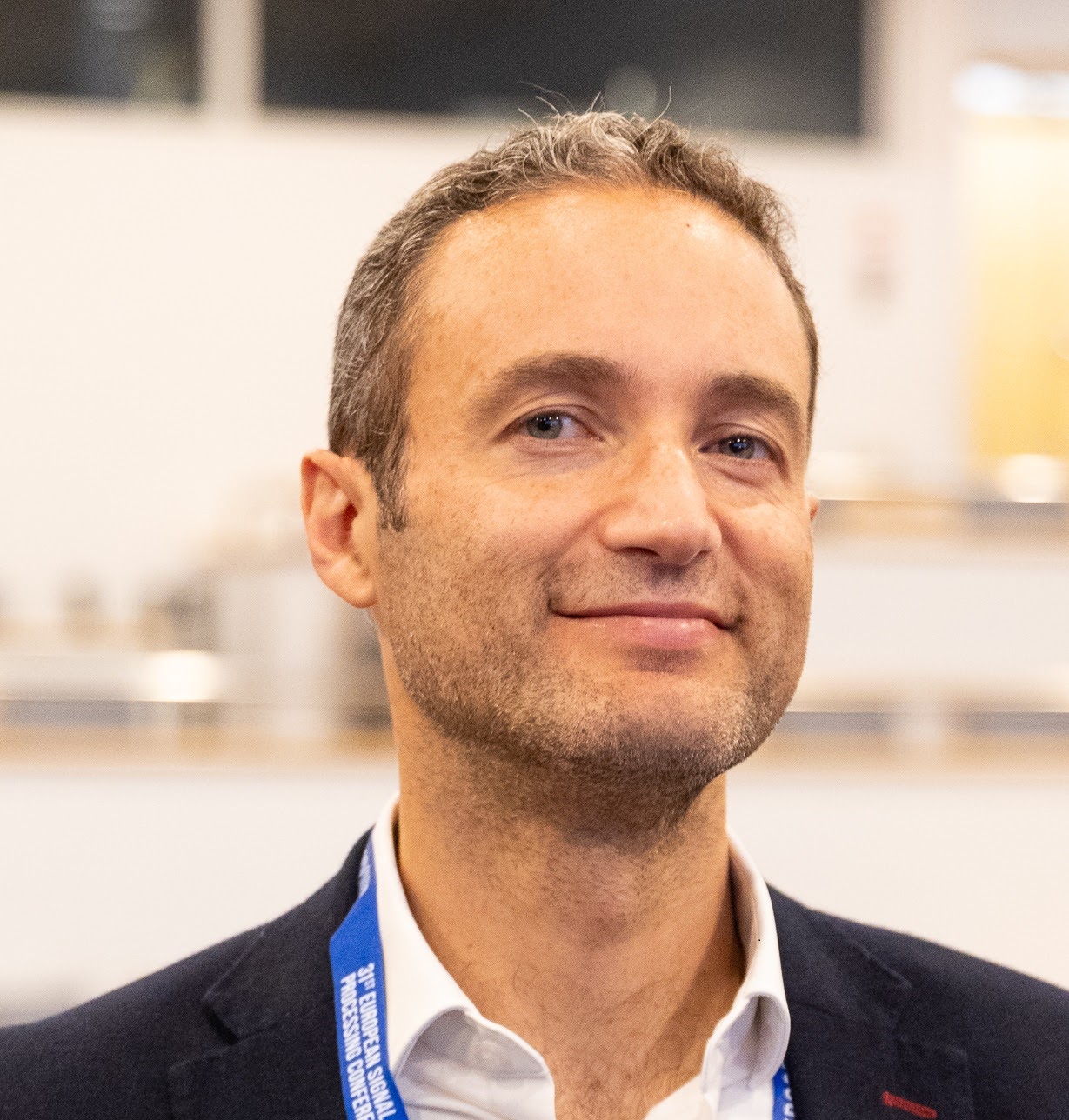}}]{Osvaldo Simeone}
is the Professor of Information Engineering at Northeastern University London, where he co-directs the Institute for  Intelligent Networked Systems (INSI).  He is also a visiting Professor with the Connectivity Section within the Department of Electronic Systems at Aalborg University. Previously, he was with the Department of Engineering of King's College London from 2017 to 2025, and with the Electrical and Computer Engineering (ECE) Department at New Jersey Institute of Technology (NJIT) from 2006 to 2017. He received an M.Sc. degree (with honors) and a Ph.D. degree in information engineering from Politecnico di Milano, Milan, Italy, in 2001 and 2005, respectively. His research interests include information theory, machine learning, wireless communications, neuromorphic computing, and quantum machine learning. Dr Simeone is a co-recipient of the 2025 IEEE ICC Best Paper Award, the 2022 IEEE Communications Society Outstanding Paper Award, the 2021 IEEE Vehicular Technology Society Jack Neubauer Memorial Award, the 2019 IEEE Communication Society Best Tutorial Paper Award, the 2018 IEEE Signal Processing Best Paper Award, the 2017 JCN Best Paper Award, the 2015 IEEE Communication Society Best Tutorial Paper Award and of the Best Paper Awards of IEEE SPAWC 2007 and IEEE WRECOM 2007. He was awarded an Advanced grant by the European Research Council (ERC) in 2025, an Open Fellowship by the EPSRC in 2022, and a Consolidator grant by the ERC in 2016. His research has been also supported by the U.S. National Science Foundation (NSF), the European Commission, the European Research Council (ERC), the Engineering \& Physical Sciences Research Council (EPSRC), the Vienna Science and Technology Fund, the European Space Agency, as well as by a number of industrial collaborations including with Intel Labs and InterDigital. He was the Chair of the Signal Processing for Communications and Networking Technical Committee of the IEEE Signal Processing Society in 2022, as well as of the UK \& Ireland Chapter of the IEEE Information Theory Society from 2017 to 2022. He was a Distinguished Lecturer of the IEEE Communications Society in 2021 and 2022, and he was a Distinguished Lecturer of the IEEE Information Theory Society in 2017 and 2018.   Prof. Simeone is the author of the textbooks "Machine Learning for Engineers" and "Classical and Quantum Information Theory" published by Cambridge University Press, four monographs, two edited books, and more than 200 research journal and magazine papers. He is a Fellow of the IET and IEEE.
\end{IEEEbiography}

 \begin{IEEEbiography}
[{\includegraphics[width=1in,height=1.25in,clip,keepaspectratio]{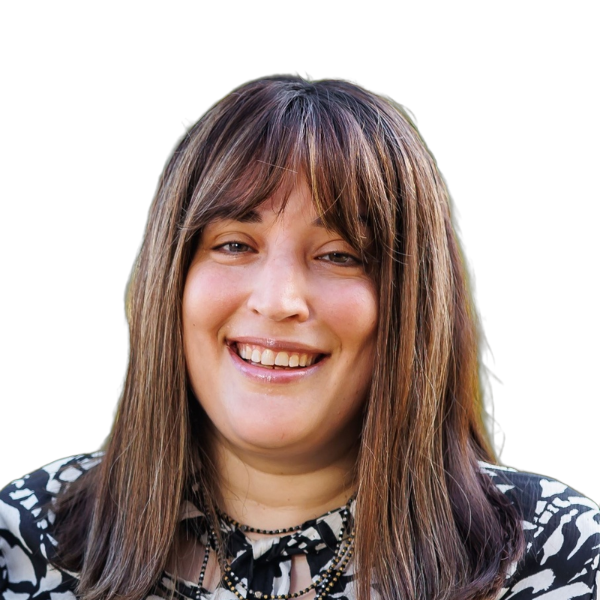}}]
{Tirza Routtenberg} (Senior Member, IEEE)
is a Professor in the School of Electrical and Computer Engineering at Ben-Gurion University of the Negev, Israel. She received the B.Sc. degree in biomedical engineering from the Technion--Israel Institute of Technology, Haifa, Israel, in 2005, and the M.Sc. and Ph.D. degrees in electrical and computer engineering from Ben-Gurion University of the Negev, Israel, in 2007 and 2012, respectively. From 2012 to 2014, she was a Postdoctoral Fellow with the School of Electrical and Computer Engineering, Cornell University, Ithaca, NY, USA. Since 2014, she has been with Ben-Gurion University of the Negev. In 2022--2023, she was a William R. Kenan, Jr. Visiting Professor for Distinguished Teaching at Princeton University, Princeton, NJ, USA.
Her research interests include statistical signal processing, estimation and detection theory, signal processing in smart grids, and graph signal processing. She served as an Associate Editor for the \emph{IEEE Transactions on Signal and Information Processing over Networks} (2019--2023) and \emph{IEEE Signal Processing Letters} (2021--2025). She has been a Subject Editor for \emph{Signal Processing} (Elsevier) since 2024. She also served as a Member of the IEEE Signal Processing Society Education Board (2023--2025) and is a member of the IEEE Signal Processing Theory and Methods Technical Committee. She was a recipient or co-recipient of Best Student Paper Awards at ICASSP 2011, CAMSAP 2013, ICASSP 2017, and SSP 2018. She received the Marc Rich Foundation Prize in 2012 and the Toronto Prize for Excellence in Research in 2021.
\end{IEEEbiography}

\begin{IEEEbiography}[{\includegraphics[width=1in,height=1.25in,clip,keepaspectratio]{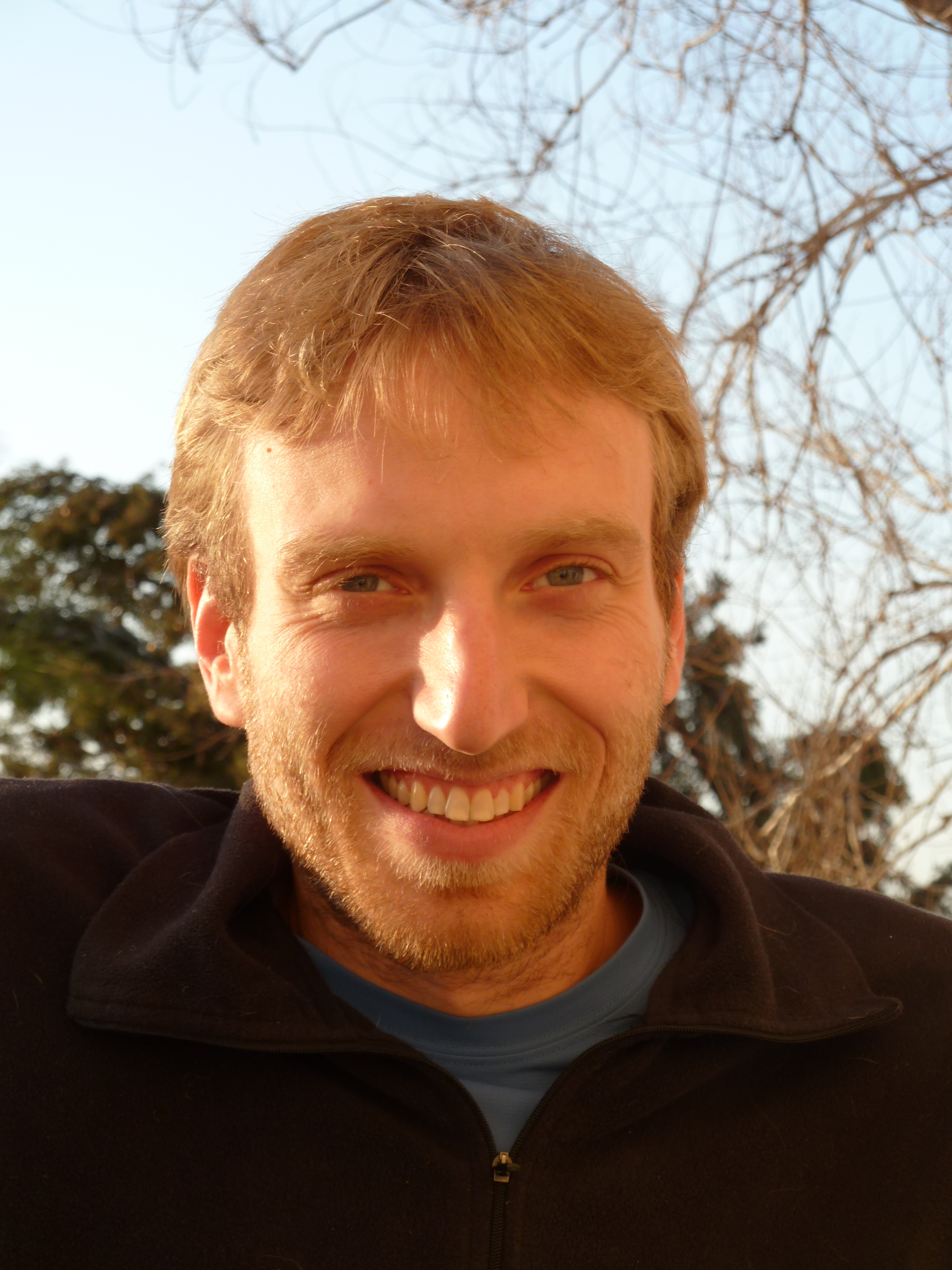}}]{Nir Shlezinger} (M’17-SM’23)  is an associate professor in the School of Electrical and Computer Engineering at Ben-Gurion University, Israel. He received his B.Sc., M.Sc., and Ph.D. degrees in 2011, 2013, and 2017, respectively, from Ben-Gurion University, Israel, all in electrical and computer engineering. From 2017 to 2019, he was a postdoctoral researcher at the Technion, and from 2019 to 2020, he was a postdoctoral researcher at the Weizmann Institute of Science, where he was awarded the FGS Prize for his research achievements. He is the recipient of the 2024 IEEE ComSoc Fred W. Ellersick Award, the 2025 IEEE ComSoc Marconi Prize, and the 2024 Krill Prize for outstanding young researchers.  His research interests include communications, information theory, signal processing, and machine learning.
\end{IEEEbiography}

\end{document}